\documentstyle{mn}      

\def\bv{$(B-V)$}
\def\vr{$(V-R)$}
\def\vi{$(V-I)$}
\def\kms{km s$^{-1}$}
\def\m100{$mag\times (100d)^{-1}$}
\def\Ha{H$\alpha$}

\def\Mb {$M_{\rm B}$}
\def\Mv {$M_{\rm V}$}
\def\Mr {$M_{\rm R}$}
\def\Mi {$M_{\rm I}$}

 \input{psfig.tex}   

\title[The peculiar SN~1991bg]
{The properties of the peculiar Type Ia SN~1991bg.\\
Analysis and discussion of two years of observations.
\thanks{Based on observations
collected at ESO-La Silla (Chile) and Asiago (Italy).}}

\author[Turatto et al.]
{M. Turatto$^{1,2}$, S. Benetti$^1$, E. Cappellaro$^2$, I.J. Danziger$^{1,3}$, 
M. Della Valle$^4$, \and C. Gouiffes$^5$, P.A. Mazzali$^3$ and 
F. Patat$^{1,4}$ 
\\
$^1$European Southern Observatory, Karl-Schwarzschild-Strasse 2, D-8046 
Garching bei M\"unchen, Germany\\
$^2$Osservatorio Astronomico di Padova, vicolo dell'Osservatorio 5,
I-35122 Padova, Italy\\
$^3$ Osservatorio Astronomico di Trieste, via G.B. Tiepolo 11,
I-34131 Trieste, Italy\\
$^4$Dipartimento di Astronomia, Universit\'a di Padova, vicolo 
dell'Osservatorio 5, I-35122 Padova, Italy\\
$^5$ DAPNIA Sap/C.E. Saclay, F-91191, Gif sur Yvette Cedex, France\\
}

\date{Accepted ...Received ... ; in original form ....}

\begin{document}

\maketitle

\begin{abstract}

Observations of the peculiar type Ia SN~1991bg in NGC~4374
collected at ESO--La Silla and Asiago are presented and discussed.
The photometric coverage extends for 530 days and the spectroscopy for
the first seven months after the explosion. The broad--band light
curves in the early months have a narrower peak and a luminosity
decline faster than other SNIa (14.6 and 11.7 \m100 in B and V
respectively). The R and I light curves do not show the
secondary peak typical of normal SNIa.  The SN is intrinsically very
red (\bv$_{max}=0.74$) and faint ($B_{max}=-16.54$).  The light curves
flatten with age but remain significantly steeper (2.0 and 2.7 \m100
in B and V between 70 and 200 day) than the average.  Consequently the
$uvoir$ bolometric light curve of SN~1991bg is fainter with a
steeper decline than that of the normal SNIa (e.g. 1992A). 
This object enhances the correlation which exists between
the peak luminosity of SNIa, the decline rate and the kinetic energy.

Peculiarities are evident in the spectra at various phases.  The
continuum at maximum is very red and the photospheric expansion
velocity extremely low. There are a number of unusual spectral
features, in particular a broad absorption between 4200 and 4500 \AA\
which is attributed to TiII and the appearance, as early as one month
after maximum, of nebular emission of possibly [CoIII]
$\lambda5890-5908$. Nevertheless, contrary to previous claims
(Ruiz-Lapuente et al 1993), the spectral evolution retains a general
resemblance to that of other SNIa until the latest available
observation (day 203).  At this epoch one sees the typical emission
features of SNIa at late times although they are significantly
narrower (FWHM$\sim2300$ \kms).  This facilitates the identification
of most lines with forbidden emission of [FeII], [FeIII] and [CoIII].
The emission feature centered at about $\lambda$6590 is difficult to
reconcile with the previous identification as \Ha, unless asymmetries
in the ejecta or ad hoc binary configurations are invoked.

This work suggests that the explosion energy was probably a factor 3
to 5 lower than in normal SNIa. Whether this resulted from an
explosion of a sub--Chandrasekhar mass WD is not unambiguously
established. 

\end{abstract}


\section{Introduction} \label{intr}

Supernovae of type Ia have been considered for decades and are still 
considered by some to be
almost perfect distance indicators. Actually, some scatter in their
photometric properties was suggested as early as the late sixties
\cite{psko1,psko2,bcr}, but it has been mostly attributed to photometric
errors (e.g. Sandage \& Tammann 1993).

However, in recent years the discovery of two extreme SNIa, 1991T and
1991bg, challenged the standard candle scenario for SNIa, 1991T being
brighter than the average SNIa \cite{ruiz_91t,phil_91t,mazz95} and
1991bg significantly fainter \cite{fili,leib}. The two SNe also showed
distinctive spectral peculiarities which, however, might have gone
unnoticed if only sparse observations had been gathered.  Therefore, a
question arises: are the samples of SNIa used as standard candles
heavily contaminated by such extreme objects and how much such
contamination affects the derived values of H$_0$ and q$_o$?  
The question has been
further complicated by the recognition that even SNIa with very
similar spectra can show a significant spread in absolute
magnitude \cite{pata}, which may correlate with the photometric
evolution and even with the morphological type of the parent galaxy
\cite{hamuh0}.

Supernovae of different intrinsic brightness have different
probabilities of being detected, hence it is expected that the
intrinsic fraction of faint SNe is higher than the actual percentage
of discoveries. This has implications for the chemical evolution
of the galaxies, since SNIa are thought to be a primary source of
Fe-peak elements.

The goal of this paper is to present and discuss new photometric and
spectroscopic observations of SN~1991bg describing the evolution of
this object until the late nebular stages and to highlight peculiarities
and similarities compared with other SNIa. Some quantitative indications on
the precursor star and the total amount of synthesized material will
also be given. 

\section{Observations and Reductions} \label{obse}

\subsection{Photometry}

SN~1991bg was discovered on 1991 Dec. 9.8 by R. Kushida \cite{kosa}
about 1 arcmin South of the nucleus of the elliptical galaxy NGC~4374,
which is located near the center of the Virgo cluster.

The observations at the ESO--La Silla and Asiago Observatories started
soon after the discovery and continued for the next seven months,
until the Virgo cluster disappeared behind the Sun. 
Eventually the SN was detected 530 days after maximum.

The broad band B, V and R photometry (together with two I-band 
observations) are listed in Tab.~\ref{phot}. Depending on the
brightness of the object and on instrument schedule, several
different telescopes were used. On photometric nights sequences of
standard stars \cite{land} were observed which allowed the
determination of the color equations of the photometric systems. These
were in good agreement with the determinations by other authors. For
calibrating the non--photometric nights, a local sequence was
established around the galaxy. For the stars in common
with the sequence of Leibundgut et al. \shortcite{leib} and Filippenko
et al. \shortcite{fili}, our measurements agree to within a few hundredths 
of a magnitude with those of the above authors.

\begin{figure*}
\psfig{figure=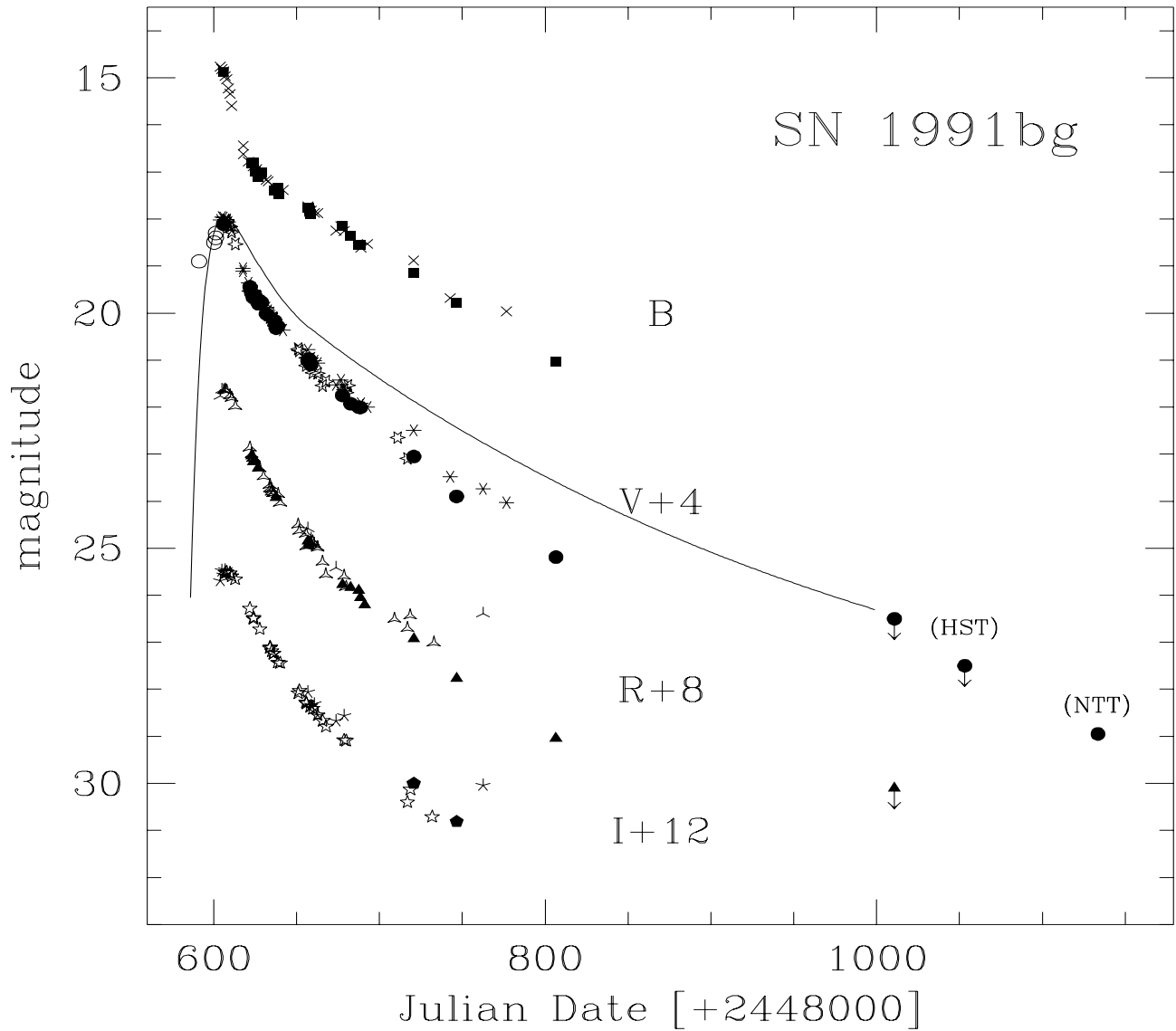,width=19cm,height=17cm}
\caption{The light curves of SN~1991bg in the BVRI bands. Ordinates are
B magnitudes while all others bands have been shifted down as
indicated.  The solid line is the V light curve of SN~1992A
(Wheeler \& Benetti 1995) adjusted to match the V maximum of
1991bg. Filled symbols are the observations presented here, starred
symbols are from Filippenko et al. (1992), skeletal symbols are from
Leibundgut et al. (1993) and open symbols from IAU Circulars. }
\label{lc}
\end{figure*}

Until 200d the photometric measurements were performed with the
ROMAFOT package, whose utilization in the context of SNe has already
been discussed (cfr. Turatto et al. 1993). The errors in the SN
magnitudes depend on the contrast of the star against the galaxy
background and were estimated to range from 0.05 at early epochs up to
0.2 mag at seven months.  In Fig.~\ref{lc} our measurements are
plotted together with those by Filippenko et al. \shortcite{fili} and
Leibundgut et al. \shortcite{leib}. It is evident that while at early
epochs the agreement between the different sources is good, it becomes
poor as the SN fades. In particular, at about 150 days our estimates
appears 0.5 mag fainter than those of Leibundgut et
al. \shortcite{leib}. Since the zero point is the same (the local
sequences coincide), the inconsistency has to be attributed to the
measuring technique. Both Leibundgut et al. and Filippenko et
al. derived the SN magnitudes by means of aperture photometry but,
while the former authors measured the
 background  in an annulus around
the SN, the latter ones, exploiting the symmetric surface brightness
profile of the elliptical parent galaxy, subtracted the image of the
galaxy rotated by $180^\circ$ around the major axis. 
We have tested the measurements of 
the SN magnitude with both these methods: already at epochs earlier than
$<100$ days the subtraction of annuli produced results dependent on
the radii of the annuli, while the subtraction of the rotated galaxy
gave results within a few hundredths of magnitude of our ROMAFOT
determination.


Based on the first, unsuccessful attempt to recover the SN, in January
1993, upper limits for the SN magnitude were derived by placing
increasingly fainter artificial stars at the precise position of the
SN relative to nearby stars. The 
result was $V \ge 22.5$ and $R \ge 22.1$.  A somewhat fainter upper limit was
obtained using an image taken with the WFC of HST on 6 March, 1993 in
the F555W passband (very close to the standard V) and retrieved from
the HST Archive. The exposure was not very deep (300 s) and the frame
was processed with the Routine Science Data Processing pipeline. Using
both the internal calibration of the frame and a faint comparison
sequence calibrated with our deep ground--based observations, we found
V$\ge 23.5$ for the SN at this epoch.  

\begin{figure}  
\psfig{figure=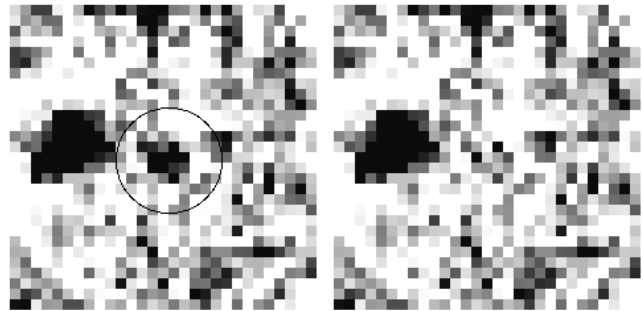,width=9cm}
\caption{Co-added V image of SN 1991bg in NGC~4374 obtained on May 26,
1993 (i.e. 530 days after maximum) at the NTT (+EMMI) 
(total exposure 170 min, seeing 1.15
arcsecs). The left panel shows the co-added original frame once the
smooth  background of the parent galaxy has been subtracted 
($\mu_V\sim21.4$ mag/arcsec$^{-2}$). The
expected position of the SN has been encircled. The right panel shows
the same area after the subtraction of the SN.
}
\label{may93}
\end{figure}

Finally, a series of 17 exposures (total observing time 170 min) was 
obtained on May 26, 1993, i.e. 530 days after B maximum.  The frames were
flat fielded, aligned and co-added into a single exposure in order to
improve the signal-to-noise ratio.  Although the seeing was not
exceptional (1.15 arcsec) a stellar image was found within 0.1 arcsec
from the expected position (Fig~\ref{may93}). In this case, the
SN magnitude was also measured using SNOOPY, a new 
software especially designed for
SN photometry by one of us (F.P.). As for ROMAFOT, in SNOOPY the
magnitude is derived by means of the point-spread-function-fitting technique,
but background fitting was improved and interactivity enhanced (Patat, in 
preparation). The resulting magnitude was V=24.95. The error of the measurement
was estimated by means of artificial star tests. The dispersion of the
recovered magnitudes of 50 artificial stars positioned near the SN
was r.m.s.=0.40 mag.

\begin{table*}
\caption{CCD photometry of SN 1991bg } \label{phot}
\begin{tabular}{lrrrrrl}    
\hline
  Date   & JD$^a$  &   B  &   V    &    R    &     I   & Instr. \\
\hline
 15/12/91& 8605.85 & 14.87& 14.11  &         &         &   NTT\\
 31/12/91& 8621.92 &      & 15.44  &         &         &   NTT\\
 01/01/92& 8622.84 & 16.82& 15.57  &  15.04  &         &  0.9m\\
 02/01/92& 8623.84 & 16.81& 15.62  &  15.16  &         &  0.9m\\
 02/01/92& 8623.85 &      & 15.67  &         &         &   NTT\\
 03/01/92& 8624.84 & 16.99& 15.63  &  15.14  &         &  0.9m\\
 05/01/92& 8626.84 & 17.10& 15.80   & 15.30  &         &  0.9m\\
 07/01/92& 8628.85 & 17.00& 15.77  &         &         &  0.9m\\
 10/01/92& 8631.85 &      & 16.02  &         &         &  0.9m\\
 15/01/92& 8636.67 & 17.40& 16.16  &         &         &   NTT\\
 16/01/92& 8637.67 & 17.38& 16.32  &  15.92  &         &   NTT\\
 17/01/92& 8638.70 & 17.34& 16.28  &         &         &   NTT\\
 18/01/92& 8639.60 & 17.47&        &         &         &   NTT\\
 04/02/92& 8656.86 & 17.75& 17.01  &  16.85  &         &   3.6m\\
 05/02/92& 8657.61 & 17.75& 16.98  &  16.94  &         &   1.8m\\
 05/02/92& 8657.88 & 17.81& 17.03  &  16.88  &         &   3.6m\\
 06/02/92& 8658.53 & 17.90& 17.10  &  16.90  &         &   1.8m\\
 25/02/92& 8677.67 & 18.15& 17.75  &  17.77  &         &   NTT\\
 01/03/92& 8682.75 & 18.35& 17.93  &  17.84  &         &   2.2m\\
 06/03/92& 8687.55 & 18.55& 18.00  &  17.90  &         &   1.8m\\
 07/03/92& 8688.54 & 18.55& 18.01  &  18.05  &         &   1.8m\\
 10/03/92& 8691.50 &      &        &  18.20  &         &   NTT\\
 08/04/92& 8720.75 & 19.14& 19.05  &  18.93  &  18.00  &   2.2m\\
 04/05/92& 8746.60 & 19.78& 19.90  &  19.77  &  18.81  &   3.6m\\
 03/07/92& 8806.52 & 21.03& 21.19  &  21.04  &         &   3.6m\\
 23/01/93& 9010.81 &      &$>22.50$&$>22.10$ &         &   3.6m\\
 06/03/93& 9053.24 &      &$>23.50$&         &         &   HST \\
 26/05/93& 9133.57 &      & 24.95  &         &         &   NTT\\
\hline
\end{tabular}

$a)$ +2440000\\
NTT=New Technology Telescope+EMMI,\\
3.6m=ESO~3.6m+EFOSC,\\
 2.2m=ESO/MPI~2.2m+EFOSC2,\\
1.8m=Asiago 1.8m+CCD camera, \\
0.9m=Dutch 0.9m+CCD camera,\\
HST=HST+WFC+F555W\\
\end{table*}

\subsection{Spectroscopy}

The journal of the spectroscopic observations is given in
Tab.~\ref{spec}, where the instrumentation and the spectral ranges
are indicated.  The spectra were calibrated in wavelength with
adjacent spectra of comparison lamps and flux calibrated with
spectrophotometric standard stars observed in the same
nights. Different exposures obtained during the same night (or in
consecutive nights at late epochs) with the same configuration have
been co-added in order to increase the signal--to--noise ratio while those
obtained with different gratings or grisms have been merged. When the
SN broad-band photometry was available on the same night, the absolute
flux calibrations of the spectra were checked. In most cases
corrections were not needed, but if necessary, corrections of the order
of few tenths of a magnitude were applied.

During the observations particular care was given to the accurate
positioning of the slit, whose width was typically between 1.5 and 2 arcsecs. 
When the spectra were obtained at large zenith distances the
slit was aligned along the parallactic angle, otherwise a second
exposure was taken with a wider slit (5--10 arcsecs) in order to
ensure that all the light from the object entered the slit.  The
spectral resolution was typically 10\AA\/ at the NTT, 12\AA\/ at the
2.2m telescope, 20\AA\/ at the 3.6m and 22\AA\/ at the 1.8m. Composite
spectra obtained by merging data from different equipment or
configurations (e.g. grisms or gratings) may thus have different resolutions 
at different wavelengths. For instance, the spectrum
at t=+117 d,  has a resolution of about 12\AA\/ between 4470 and
8420\AA\/ and about 30\AA\/ outside this range (cfr. Fig~\ref{sp_ev}).

\begin{table}
\caption{Journal of the spectroscopic observations of SN~1991bg } \label{spec}
\begin{tabular}{lccccccl}    
\hline
  Date & phase$^a$ &  JD$^b$ &   Wavelength & Instrum. \\
         & (days)&         &   range (\AA)     &          \\
\hline
 14/12/91& 0.7	& 8604.65 & 3650-9250		& 1.8m	\\
 15/12/91& 1.6	& 8605.60 & 4030-9200		& 1.8m	\\
 15/12/91& 1.9	& 8605.85 & 3880-8120		& NTT	\\
 29/12/91& 15.6	& 8619.57 & 4170-8450		& 1.8m	\\
 31/12/91& 17.9	& 8621.88 & 4170-9800		& NTT	\\
 02/01/92& 19.5	& 8623.85 & 4170-7860		& NTT	\\
 14/01/92& 31.5	& 8635.54 & 3850-9420		& 1.8m	\\
 15/01/92& 32.5	& 8636.54 & 6850-11000		& 1.8m	\\
 16/01/92& 33.7	& 8637.67 & 4160-7950		& NTT	\\
 28/01/92& 45.6	& 8649.56 & 3560-9040		& 1.8m	\\
 01/02/92& 49.8	& 8653.83 & 3560-7100		& 2.2m	\\
 04/02/92& 52.7	& 8656.65 & 3560-9200		& 1.8m	\\
 04/02/92& 52.9	& 8656.87 & 3600-6720		& 3.6m	\\
 13/02/92& 61.6	& 8665.56 & 4050-8410		& 1.8m	\\
 25/02/92& 73.7	& 8677.67 & 5400-6500		& NTT	\\
 01/03/92& 78.8	& 8682.79 & 4470-8200		& 2.2m	\\
 10/03/92& 87.5	& 8691.50 & 4200-10000		& NTT	\\
 08/04/92& 116.7& 8720.71 & 3560-9100		& 2.2m	\\
 04/05/92& 142.6& 8746.60 & 3730-6900		& 3.6m	\\
 03/07/92& 202.5& 8806.52 & 3770-6900		& 3.6m	\\
\hline
\end{tabular}

$a)$ from B maximum (J.D. 2448604.0)\\
$b)$ +2440000\\
NTT=New Technology Telescope+EMMI,\\
3.6m=ESO~3.6m+EFOSC,\\
2.2m=ESO/MPI~2.2m+EFOSC2,\\
1.8m=Asiago~1.8m+B\&C
\end{table}

\section{Light curves}  \label{licu}

The light curves of SN~1991bg in the first months after maximum,
already discussed by Filippenko et al. \shortcite{fili} and Leibundgut
et al. \shortcite{leib}, are updated in Fig.~\ref{lc} with the
inclusion of our new data. It is confirmed that the photometric
evolution of SN~1991bg differs from that of normal SNIa: the maximum
peak is narrower and the luminosity decline faster.  Premaximum
observations are scanty and quite uncertain, therefore little can be
said about the rise to maximum light. Our best estimates of the epochs
and magnitudes at maximum are reported in Tab.~\ref{data} and are in
good agreement with previous estimates.

\begin{figure} 
\psfig{figure=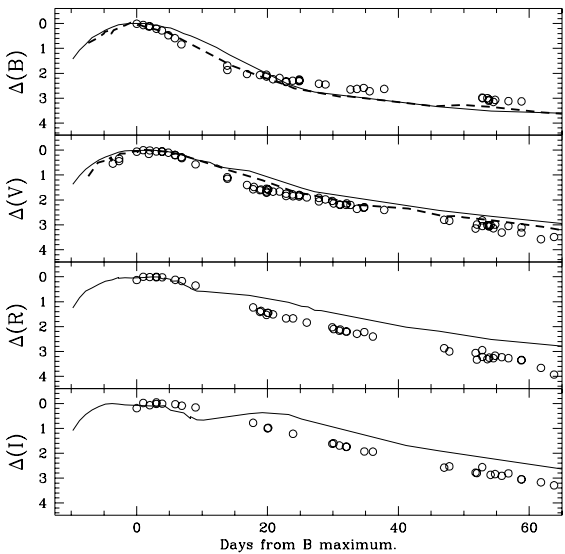,width=8.5cm}
\caption{Comparison between the early light curves of SN~1991bg
(symbols), SN~1994D (solid lines, Patat et al. 1994) and SN~1986G
(short dashed line, Cristiani et al. 1992).  In the four panels, zero
in the abscissa corresponds the epoch of the B maximum while in the 
ordinates it is the magnitude at maximum in the specific photometric
band.}
\label{confro}
\end{figure}

The light curves of SN~1991bg in the first 2 months after maximum are
compared in Fig.~\ref{confro} to those of the normal SNIa 1994D and of
the somewhat peculiar SNIa 1986G.  The general behavior of SNIa,
showing a slower luminosity decline in the red than in the blue, holds
also for SN~1991bg which however shows the steepest initial luminosity
decline rate, namely 14.6 and 11.7 \m100 in B and V respectively, to
be compared with 12.5 and 7.2 \m100 measured in SN 1994D
\cite{pata}. The secondary maximum, which is quite pronounced in the I
light curve of SN~1994D, is absent in SN~1991bg.

The light curve evolution of SN~1986G is also faster than that of
SN~1994D (12.8 
and 8.6 \m100 in B and V respectively) albeit not as fast as SN~1991bg
(cfr. Fig.~\ref{confro}).  Unfortunately the R and I light curves of
SN~1986G are not available, and so a comparison with SN~1991bg, which
departs strongly from the standard SNe~Ia in these bands, is not
possible but neither SN shows the secondary maxima in the near--IR
typical of normal SNIa \cite{frog,port}.


As early as 15 days after maximum, the B light curve of SN~1991bg
begins a slower decline, while at increasingly redder wavelengths the
change of slope occurs later and is less evident. Therefore, the value
of $\Delta(m)_1$ given in Tab.~\ref{data} for the B band corresponds
to the parameter $\delta m_{15}(B)$ defined by Phillips
\shortcite{phil}.  Our determination is in good agreement with his
estimate even after the inclusion of our new photometry.

\subsection{The photometric evolution}  \label{pe}

\begin{table*}
\caption{Basic photometric data of SN~1991bg} \label{data}
\begin{tabular}{lccccc}
 &	B	&	V	&	R	&   I	& bol\\  
\cline{2-6}
&       &               &               &       &\\
\multicolumn{6}{c}{MAXIMA}\\
\hline
J.D.(+2448000)& $\sim 604.0$ &$605.5\pm1.0$ &$606.0\pm1.0$
&$608.0\pm1.0$ & $605.0\pm1.0$\\
 mag.  & $\sim 14.75\pm0.1$ &$13.96\pm0.05$ &$13.63\pm0.05$ & $13.50\pm0.05$&\\
abs.mag$^*$&$-16.54\pm0.32$ &$-17.28\pm0.31$ & $-17.58\pm0.31$
& $-17.68\pm0.31$ & $-16.50\pm0.34$ \\ 
$\log L$ [erg s$^{-1}$] & & & & & ~~$42.20\pm0.14$ \\
\\ 
                &       &               &               &       &\\
\multicolumn{6}{c}{FADING RATES [\m100] (phase range [days])} \\
\hline
early	&14.6(3-14)	& 11.7(6-14)&  9.4(6-24)& 7.3(9-24)& 9.1(4-20)\\
intermediate &2.6(17-59) & 4.2(17-64)& 4.6(30-59)& 4.8(30-64)& 3.6(30-70)\\
late &2.0(70-203) & 2.7(70-203)&  2.6(70-203)& 2.7(75-143)& 2.6(70-203)\\
very late         &       	& 1.2(203-530)&               &  &\\
                        &       &               &               &  &\\
\multicolumn{6}{c}{BENDING POINTS}\\
\hline
epoch$_1$ [days]	&15	&	17	&24	&	28 & 22\\
$\Delta(m)_1$ [mag]	&1.95	&	1.44	&1.77	&	1.50 & 1.70\\
			&	&		&&&\\
epoch$_2$ [days]	&60     &       60      &68      &       65 & 70\\
$\Delta(m)_2$ [mag]     &3.25   &       3.34   &3.87    &       3.30 & 3.50\\
                        &       &               &&&\\
\end{tabular}

\begin{tabular}{lcccl}
COLORS	& \bv	&$(V-R)$	&$(V-I)$	\\
\hline
day 0 (B max.)		& 0.74	&0.25		&0.33		\\
day 15--17		&1.45	&0.60		&1.25		\\
\hline
\end{tabular}

\begin{tabular}{lccccc}

\phantom{abs.mag$^*$} & \phantom{$-16.54\pm0.32$} & \phantom{$-17.28\pm0.31$} 
& \phantom{$-17.58\pm0.31$} & \phantom{$-17.68\pm0.31$} & 
\phantom{$-16.50\pm0.34$} \\ 
\multicolumn{2}{l}{(*) $E(B-V)=0.05$, $\mu=31.09$} & &&&\\
\end{tabular}

\end{table*}

Previous analyses have shown that the light curve of SN~1991bg can be
 well approximated by 3 segments spanning from day 4 to 20, 20 to 60
 and 60 to 200, respectively.  The inclusion of our new data changes
 the positions of the bending points slightly with respect to earlier
 determinations \cite{fili,leib}.  The epochs of the bending points
 and the slopes (in units of \m100) of the three segments are given in
 Tab.~\ref{data}.  As we will argue later, the fast evolution of the
 early light curve are indicative of a small ejecta mass.

The new observations of Tab.~\ref{phot} are especially important since
they constrain the late time photometric behavior of SN~1991bg.  Two
months after maximum, the second bending point in the light curves marks 
the beginning of the radioactive tail which then remains linear until at
least day 200 (Tab.~\ref{data}). Again the luminosity decline rate of
SN~1991bg in this phase is significantly faster than those of all
other well-observed SNIa, which typically have a rate of 1.5 \m100
(cfr. Turatto et al. 1990, Suntzeff 1995).

Finally, the observation 530 days after maximum 
indicates that after day 200 the luminosity decline was significantly
slower, 1.2 \m100 based only on the two last observations.  Due to the
300 day gap in the observations, it is impossible to determine if the
change was gradual or if a new bending point characterizes the late
light curve.  We note, however, that the observed decline rate is
still steeper than that of $^{56}$Co to $^{56}$Fe decay (0.98 \m100).


\subsection{The absolute magnitude at maximum}  \label{abma}

NGC 4374, the parent galaxy of SN~1991bg, lies in the core of the
Virgo cluster. This makes it possible to compare directly the
apparent magnitude of SN~1991bg with that of other SNIa in the
cluster.  A list of the Virgo SNIa with ``normal'' spectra and
reliable photometry was given by Patat et al. \shortcite{pata} 
(their Tab.~4): the average apparent magnitude is $B_{max}=12.22\pm0.44$,
to be compared with $B_{max}\sim 14.75$ for SN~1991bg.  Two other SNIa
have been discovered in NGC 4374, SN~1957B and 1980I (cfr. Turatto et
al.  1994). Both SNe show apparent magnitudes in agreement with that of
normal SNIa in Virgo. Therefore, if no significant reddening was
present, SN~1991bg was at maximum 2.5 magnitudes fainter in the B band
than normal SNIa.

Strong reddening is not expected, since the parent galaxy is
elliptical. Actually a dust lane, elongated in the E--W direction,
crosses the galaxy nucleus (cfr. Filippenko et al. 1992, Leibundgut et
al. 1993) but the SN is located about 1 arcmin to the south of the
nucleus and hence no significant contamination is expected.  In fact,
a careful examination of our best S/N spectra (cfr. Sect. \ref{spev})
did not show signs of narrow interstellar NaI~D at the expected
position (5913\AA).  In a few spectra a weak absorption might be
present at about 5890 \AA\ that can be due to interstellar NaI~D
within the Galaxy ($A_B=0.13$, Burstein \& Heiles 1984). Indirect, but
independent, support for low reddening comes from the spectral
modeling at early and late epochs by Mazzali et
al. \shortcite{91bg:mod}.  Those models require low photospheric
temperatures. This, together with the fact that all 3 colours, $B-V$,
$V-I$, $V-R$ are redder than normal SN~Ia immediately after maximum
but return to normal colours 30 days past maximum, suggests that
neither internal dust formation nor line blanketing due to additional
abnormally strong lines are the cause of the red colours.  In
agreement with other papers on this SN, we will adopt in the following
E(B-V)$\sim0.05\pm 0.02$ as the total reddening suffered by
SN~1991bg. Given the low reddening, it is clear that SN~1991bg
was {\em intrinsically} faint and red.

For homogeneity with  Patat et al. 1995 and  Vaughan et al.
\shortcite{vaug} who studied large samples of SNIa, we adopt for 
NGC 4374 the distance derived using the SBF method
($\mu=31.09\pm0.30$, Ciardullo et al.  1993).
Taking into account the
(small) reddening, we obtain for SN~1991bg the following absolute magnitudes:
\Mb$=-16.54\pm0.32$, \Mv$=-17.28\pm0.31$, \Mr$=-17.58\pm0.31$ and 
\Mi$=-17.68\pm0.31$. 
These values are significantly fainter and redder than the mean,
\Mb$=-18.64\pm0.05+5 log(H_o/85)$ and \Mv$=-18.63\pm0.06+5 log(H_o/85)$,
found by Vaughan et al. \shortcite{vaug} for their {\em ridge--line}
SNIa, i.e. SNIa which suffered little extinction.
SN~1991bg is even more so compared to the average SNIa magnitudes derived with 
the Cepheid distance calibration (\Mb$=-19.65\pm0.13$ and \Mv$=-19.60\pm0.11$,
Saha et al. 1995). 

At present, it is widely accepted that some degree of heterogeneity is
present within the class of SNIa (e.g. Patat et al. 1995, Hamuy et
al. 1995). This may be partially described by a relation between the
peak luminosity and the initial decline rate \cite{psko1,phil,hamuh0}. 
However, even in this context the low luminosity of SN~1991bg is 
exceptional, since this SN does not lie on the extension of the peak 
luminosity--decline rate relation observed for the slowly declining objects 
\cite{hamuh0}. 

\begin{figure} 
\psfig{figure=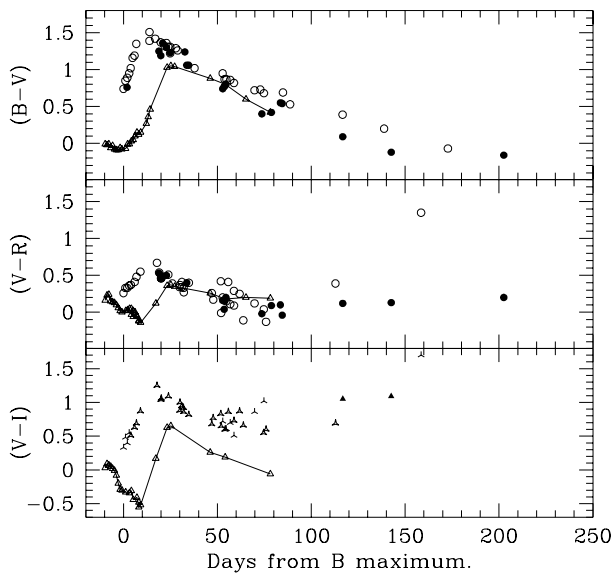,width=8.5cm}
\caption{The color curves of SN~1991bg compared to those  
of SN~1994D (connected by solid lines) which has a 
reddening similar to SN~1991bg (Patat et al. 1995). Filled symbols are
observations from Table~1, open symbols are data from
literature. Note the scatter at late epochs.}
\label{colo}
\end{figure}

\subsection{Color curves}  \label{col}
The evolution in three different colors is shown in Fig.~\ref{colo}
in which the comparison is made with the well-studied SN~1994D which had
comparable reddening to 1991bg and whose color evolution was similar to 1992A 
\cite{pata}.

The two SNe show significantly different color evolutions.  At maximum,
SN~1991bg had \bv$\sim0.75$, i.e. it was $\sim 0.8$ mag redder than
SN~1994D. Were the color differences due to reddening for
SN~1991bg, which we excluded (cfr Sect.~\ref{abma}), this would indicate
$A_B=3.2$ yielding \Mb$=-19.54$. With such values SN 1991bg would be
the brightest SNIa in the samples studied by Vaughan et
al. \shortcite{vaug} and Phillips \shortcite{phil}.  At the time of maximum, 
SN~1994D was still evolving to the blue, with minima in \vr\ and
\vi\ corresponding to the minima between the two peaks
of the R and I light curves \cite{pata}. On the contrary, SN~1991bg at the 
same epoch was already rapidly evolving to the red. SN~1991bg reached the 
reddest colors \bv=1.45, \vr=0.60 and \vi=1.25) at about t=15-17d, 
corresponding to the first bending point of the B light curve, 
while in SN~1994D the maxima of the colors were reached somewhat later 
(t=25d) at \bv=1.05, \vr=0.37 and \vi=0.65.  Diversity in the color 
evolution of SNIa (especially in proximity of maximum)  have been noted by 
Maza et al. \shortcite{maza} and Patat et al. \shortcite{pata} but, to our
knowledge, no other object appears as odd as SN~1991bg.

After the color maxima, SN~1991bg turned to the blue and starting about
day 40-50 it matched the \bv\ and \vr\ curves of normal SNIa
(Fig~\ref{colo}).  The \vi\ curve, instead, never conformed
with that of 1994D, remaining always considerably redder. This gives us
an indication that the late overall light distribution of
SN~1991bg is similar to normal SNIa but for a flux excess in the I
band.  In Sect.~\ref{nebu} this is 
ascribed to the unusually strong [CaII] emission lines.

\begin{figure} 
\psfig{figure=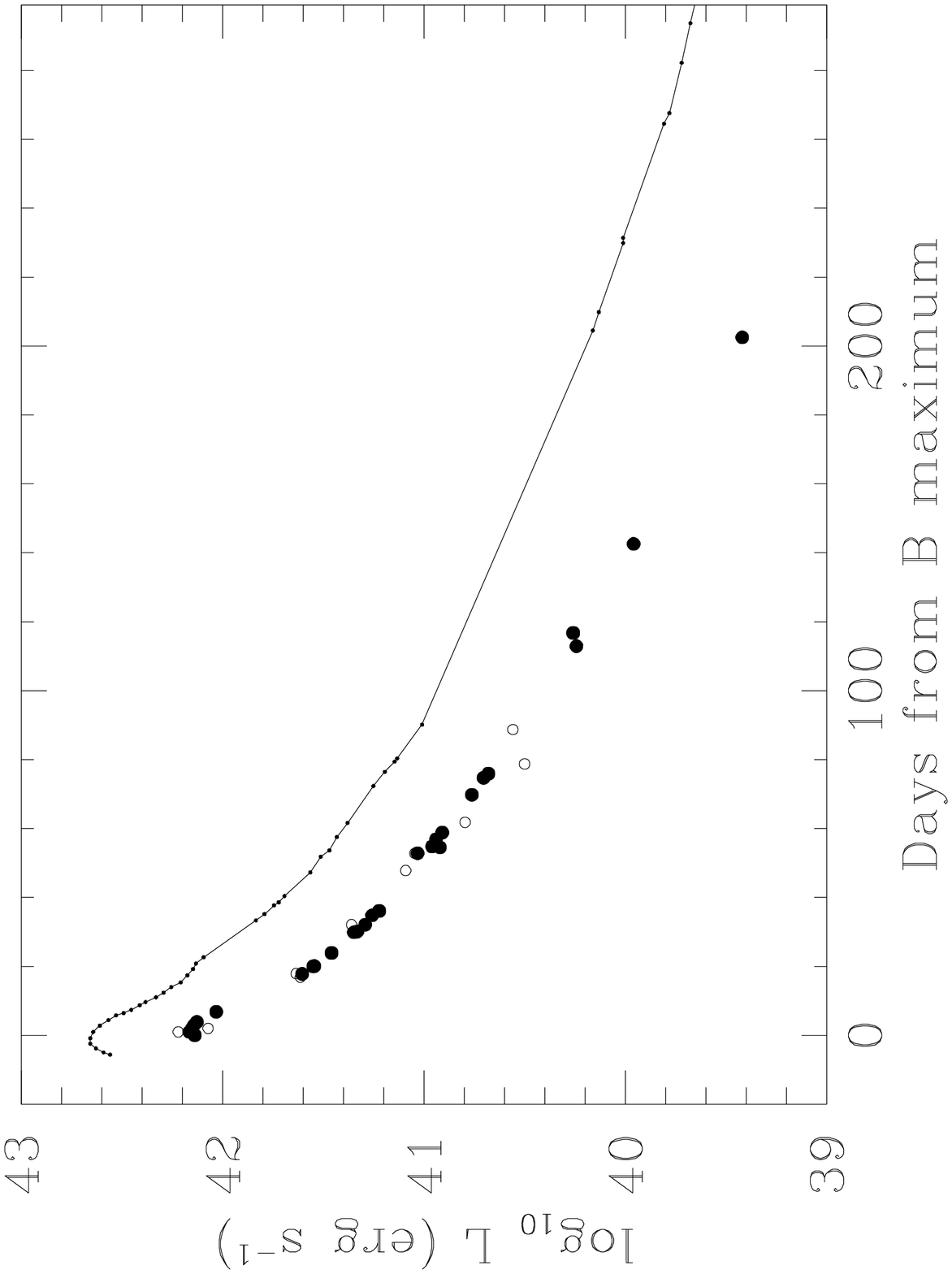,width=8.5cm,rotate=90}
\caption{Bolometric light curves of SNe 1991bg (symbols) and 1992A
(line). Filled symbols refer to BVRI photometry from this paper and
>from literature while open symbols are derived from the flux calibrated
spectra of Table~2 with wide wavelength range. For the computation
of the last point (day +203) the I photometry has been linearly
extrapolated form the previous epochs.
For SN~1991bg we have used a bolometric correction of 0.1 dex constant
with time (see text for a discussion of the relative luminosity), 
distance modulus 31.09 and reddening correction
E(B-V)=0.05.  The data for SN~1992A are from Suntzeff (1995).}
\label{bolofig}
\end{figure}

\begin{figure*} 
\psfig{figure=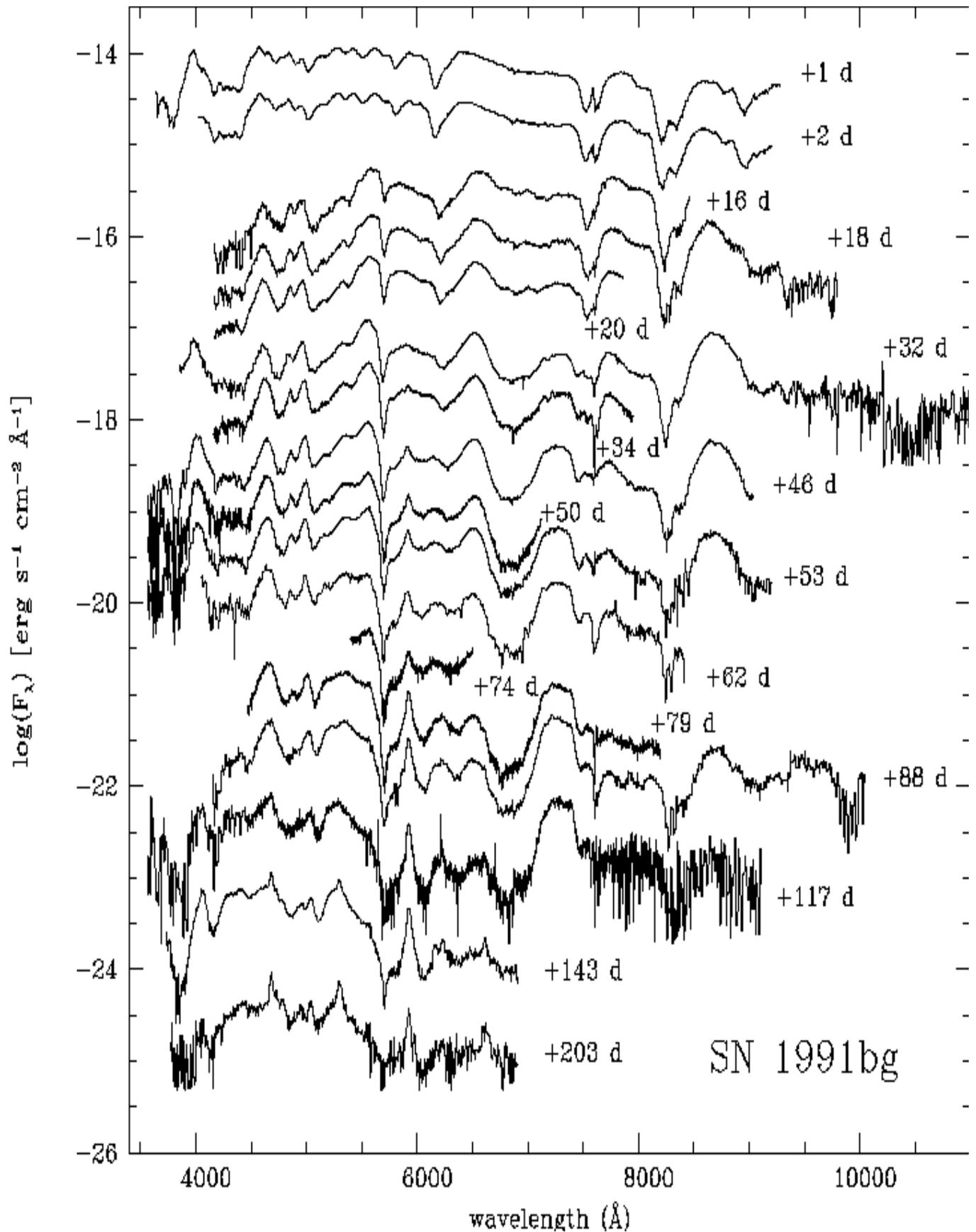,width=18cm,height=23cm}
\caption{The spectral evolution of SN 1991bg from day +1 to day +203.
The ordinate refers to the first spectrum (t=+1 d); all other spectra 
are shifted downwards by 0.45 dex with respect to the one above,
with the exception of the last three, for which the shift is 0.65 dex.}
\label{sp_ev}
\end{figure*}

\subsection{The bolometric light curve}  \label{bolo}

The optical light curves of SN~1991bg are well defined in four bands
for several months. These allowed us to estimate the {\em uvoir}
bolometric luminosity by applying reasonable corrections to the
optical data.  Suntzeff \shortcite{sunt} showed that in normal SNIa at
early phases the flux below 4000 \AA\ is less than 40\% of the total
flux and rapidly declines below 10\% already at 80 days after maximum.
Also the contribution of the IR above 9000 \AA\ is always low (less
than 15\%,  Suntzeff \shortcite{sunt}). 
In the case of SN~1991bg the dominance of the BVRI bands
is even larger: merging the (single) IUE spectrum available, taken at
B maximum \cite{ulda}, with our nearly--simultaneous optical
observations, we have estimated that the flux below 4000 \AA\ is only
$\sim 10\%$ of the total flux in the range 1900--9000 \AA.  Also the
JHK infrared observations of SN~1991bg \cite{port} indicate a low
luminosity at maximum followed by a rapid decline of 3 mag in the
first month, with no sign of a secondary maximum.  It is reasonable to
assume that in SN~1991bg the UV contribution decreased with epoch
while the near-infrared one increased up to 15\% of the total {\em
uvoir} flux analogous to SN~1992A \cite{sunt}.
This means that at all epochs about 80\% of the total flux of SN~1991bg is
emitted in the BVRI domain, corresponding to  a constant
correction of 0.10 dex.  Spectral
modeling, discussed by Mazzali et al. \shortcite{91bg:mod}, gives
results in good agreement with this assumption.

The evolution of the bolometric luminosity is compared in
Fig.~\ref{bolofig} to that of SN~1992A \cite{sunt}. Both BVRI photometry
(from this paper and literature) and spectrophotometric data have been
used. A distance modulus of 31.09 and a reddening correction
E(B-V)$=0.05$ have been adopted (cfr. Sect.~\ref{abma}). \footnote{The
maximum bolometric luminosity of SN~1991bg, $log_{10} L=42.20$, is
about 0.25 dex (i.e. a factor 1.8 in luminosity) fainter than in
Fig. 7 of Suntzeff \shortcite{sunt} who adopted a similar distance
modulus and no reddening correction.  The inconsistency is probably
due to a different correction factor.}  Thus 
SN~1991bg at maximum was about 3 times fainter than the 
spectroscopically normal SNIa
SN~1992A ($log L=42.65$) and, because of a faster luminosity
decline, the difference increased to a factor 5 on day 200.
 It should be noted that, according to Suntzeff  \shortcite{sunt} 
either the distance scale to SN~1992A is grossly in error, or this
type Ia SN
is underluminous with respect to the standard model, so the offset in peak 
L$_{Bol}$ values in Fig.~\ref{bolofig} 
does not indicate that SN~1991bg is a factor 3 fainter 
than normal SNe Ia. 

In Tab.~\ref{data} we present the main data of the bolometric light
curve. The slopes have been determined making use only of the photometric
determinations. It is evident from Fig~\ref{bolofig} that SN~1991bg
was at all epochs steeper than SN~1992A which
declined at a rate of 2.0 \m100 between day 72 and day 210.

At the last epoch of observation (day 530) only the V magnitude is
available, and thus no estimate of the bolometric luminosity is
possible. We note, however, that assuming at this epoch the same
spectral distribution as on day 203, the corresponding {\em uvoir}
bolometric luminosity is about $10^{38}$ ergs s$^{-1}$.

\section{The spectral evolution}  \label{spev}
The first available spectra of SN~1991bg were taken near 
maximum light. Figure \ref{sp_ev}, which shows the spectra listed
in Tab.~\ref{spec}, illustrates the overall spectroscopic
evolution from the time of maximum to 7 months later.

\subsection{The first two months}
\begin{figure}
\psfig{figure=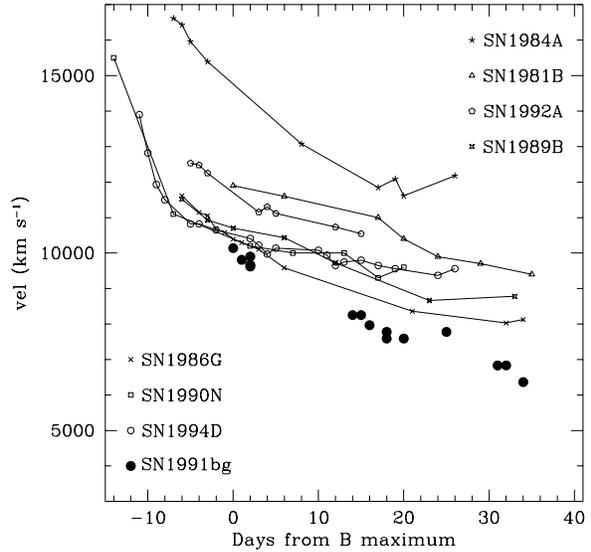,width=8.5cm}
\caption{Expansion velocities of SNIa as deduced from the minimum of the SiII
$\lambda 6355$ line. Data of SN~1991bg are from this paper and
Leibundgut et al. (1993).}  \label{exp}
\end{figure}

Although it showed the distinctive SiII absorption, the first
available spectrum, taken at maximum, appeared different from those
for typical SNIa.  In Fig.~\ref{exp} the time evolution of the
photospheric velocity deduced from SiII $\lambda 6355$ is
plotted. Clearly, the expansion velocity of SN~1991bg was quite low
(at maximum, $v_{t=0}=9720$ \kms) and declined faster.  Also, the blue
wing of the SiII doublet, though possibly blended with another weak
line, indicates a maximum expansion velocity for the SiII-rich layers
of only about 14500 \kms, to be compared to 16000 \kms\ of SN~1994D
\cite{pata}.

\begin{figure}
\psfig{figure=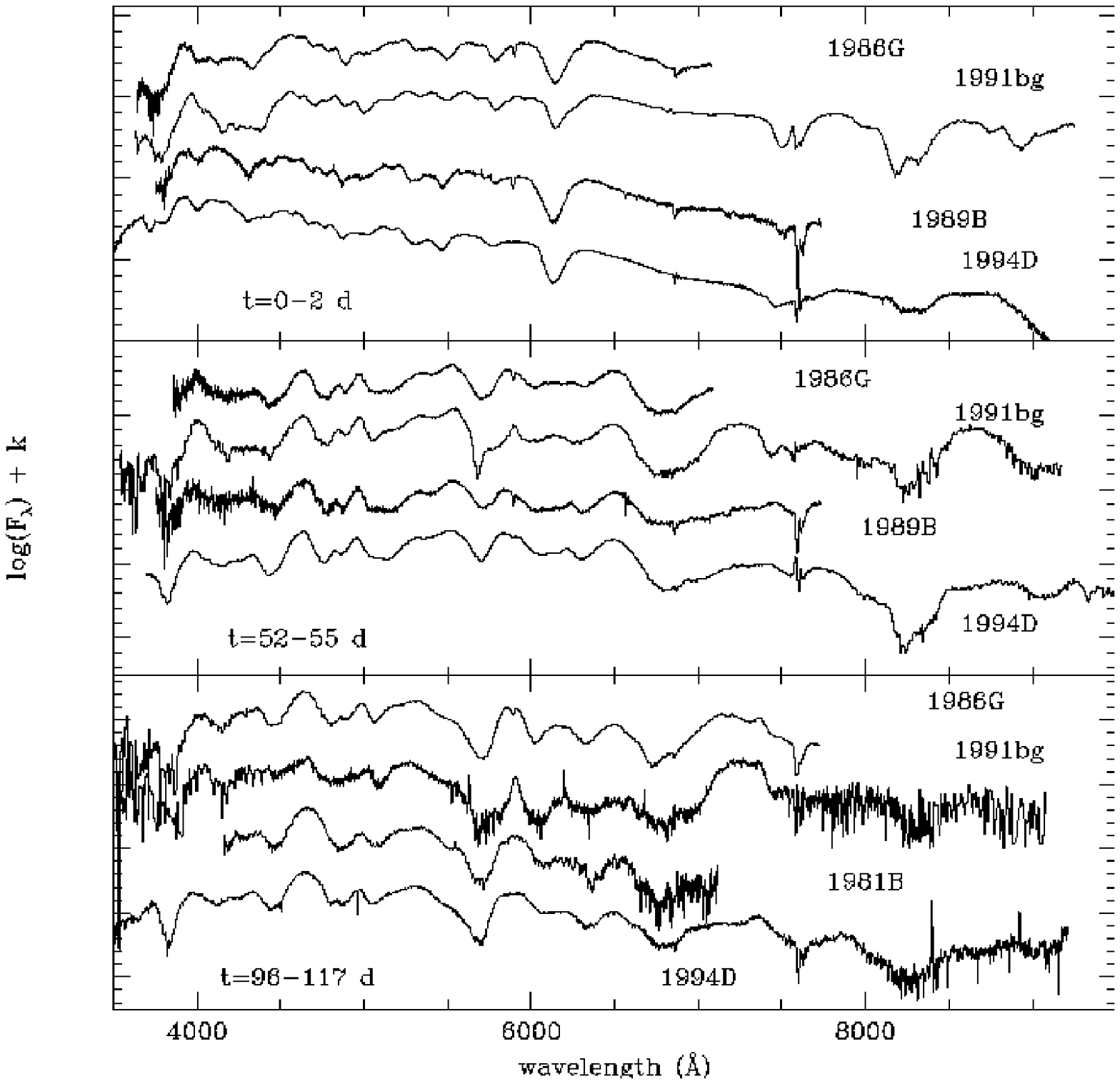,width=8.5cm,height=13cm}
\caption{Comparison of the spectra of SN 1991bg at maximum (top
panel), around 50 (middle) and 100 (lower) days past maximum
with those of other SNIa: the ``normal'' SNe 1994D and 1981B, 1989B
and the peculiar SN 1986G. The spectra of SN~1986G and SN~1989B have
been dereddened by E(B-V)=0.6 and 0.35, respectively (Phillips 1993).
The tracings have been displaced by arbitrary units and reported to the
rest frames of the parent galaxies. References are Cristiani et al.
(1992) for SN~1986G; Barbon et al. (1990) for SN~1989B; Patat et al. (1995) 
for SN~1994D (early times); Branch (1984) for SN~1981B. 
}
\label{earl}
\end{figure}

To highlight the peculiarities of SN~1991bg, in Fig.~\ref{earl} the
spectra of SN~1991bg at different epochs are compared to those of
other SNIa at similar phases. Only SNIa with low expansion velocities
 have been selected. At maximum, the slope of the
optical continuum differs from that of the normal SNIa 1994D and
1989B, once the latter is dereddened.  Instead, it resembles the slope
of SN~1986G (cfr. top panel of Fig.~\ref{earl}) if one adopts for this
SN a reddening E(B-V)=0.6 \cite{phil}.  Other similarities with
SN~1986G can be seen, in particular in the broad absorption trough
between 4200 and 4500 \AA\ which has been early on noted
\cite{ben91}. Filippenko et al \shortcite{fili} attributed this feature
to MgII 4481 \AA\/ and TiII lines that could also account for a
number of other features below 5000 \AA, while Leibundgut et
al. \shortcite{leib} identified it with FeIII. The spectrum 
synthesis analysis \cite{91bg:mod} supports the TiII identification.

As in SN~1986G, the red wing of the SiII~$\lambda6355$\AA\ absorption 
seems to be contaminated by another absorption line. After deblending, 
the latter  was measured at about 6255 \AA\/ in the parent galaxy rest frame. 
Leibundgut et al. \shortcite{leib} have proposed CII~$\lambda$6580 \AA\/
or \Ha\ from hydrogen expanding at velocities of the order of 14000 \kms as
possible identifications.
The similarity with SN~1986G extends to the absorption at about 5800
\AA\ attributed to SiII $\lambda5968$, which in both SNe
is stronger than in normal SNe~Ia. Also, the two lines of SII
near 5500\AA\ have similar intensity, while the redder one
is usually stronger  in other SNe~Ia at this epoch.

As in other SNe~Ia spectra, in the near--IR the strong P-Cygni profile 
of the CaII triplet is possibly contaminated by OI~$\lambda8446$.
The absorptions at 7600 \AA, due
to OI~$\lambda7772-7775$ and MgII~$\lambda7877-7896$, and that at 8900
\AA, due to MgII~$\lambda9217-9243$, are instead much stronger than usual.

There is a two week gap in the temporal coverage both in our
spectroscopic observations and in those of Filippenko et
al. \shortcite{fili} and Leibundgut et al. \shortcite{leib}.  When the
SN was observed again at the end of December, about 2 weeks after 
maximum, the spectrum was quite different (cfr. Fig.~\ref{sp_ev}).  
At this point the rapid post-maximum decline phase 
is over, and the SN enters what, in Sect.~\ref{licu}, we called
the {\em intermediate} phase of the light curve. The continuum is now
redder (cfr. also Fig. \ref{colo}) and emission lines, which are commonly
observed in SNIa only at later epochs, start to appear.  The
$\lambda 6355$ SiII absorption is still present, while the
corresponding emission has strengthened and has moved redward. 
[NiIII] $\lambda 6536$ may also contribute to this emission. 
On the other hand, the SiII absorption at 5810 \AA\ (5792 \AA\/ in the
galaxy rest frame) disappeared, 
probably as a result of the drop in temperature lowering the
population of the lower level which is 2 eV higher than the
corresponding level giving the SiII 6355 line. 
The CaII IR lines are very strong and the lines of OI
and MgII have decreased their intensities. Now the Ca emission
dominates over the absorption.

The strong and relatively narrow absorption feature, measured on day 18 
at about 5705 \AA\ (5687 \AA\/ in the galaxy rest
frame\footnote{Throughout this paper we have adopted for SN~1991bg the
recession velocity of NGC4374, $v=933$ \kms \cite{tully}} is quite unusual.
This line, which was also barely visible in the spectra at maximum,
will persist for about 5 months.  The line profile is strongly
asymmetric but definitely narrower than all other absorption features
in the spectrum, FWHM$\sim2500$ \kms, and constant with time.
If the feature is attributed to NaID \cite{fili}, the expansion
velocity of the absorbing layer, $v_{NaI}=10500$ \kms, is much larger
than that of SiII ($v_{SiII}=7600$ \kms) at this epoch.  Leibundgut et
al. \shortcite{leib} attributed this feature at early epochs to a
blend of SiII lines but its narrowness with respect to other SiII
lines and its persistence after SiII 6355 has faded 
makes this doubtful.  An unlikely identification 
is HeI~$\lambda5876$ expanding at about 9600 \kms, because of the low 
photospheric temperature, unless non-thermal excitation of appropriate
levels is effective. We lack independent evidence of non-thermal
effects. In any of these cases, however, the high expansion velocity 
and the narrow width 
confine the ion from which the line forms to a relatively thin outer 
layer. This would then be clear-cut evidence of stratification 
in SN~Ia ejecta. 

However the feature changes very little in velocity over the 200 days
of observation, in contrast with other absorption features. 

The line drifts from 5710 \AA\ on day 1 to 5700 \AA\ at about 50 days
past maximum, in good agreement with the measurements of Leibundgut
\shortcite{leib}. 
Thus the correct interpretation may have to await a detailed modeling
to ascertain whether radiation transfer effects which describe the
redward progression of photons in the expanding photosphere can
produce this feature. 

On day 50 all absorptions (5700 \AA\ feature excepted) 
have drifted notably to the red with respect
to the initial position, indicating that the photospheric expansion
velocity has decreased significantly. The CaII lines appear at early
phases stronger in absorption than in other SNIa, in particular the 
IR triplet; and stronger in emission at later phases, in particular 
[CaII] $\lambda 7291-7324$. Although abundance effects may play a
role, a temperature effect must be important. Apart from this peculiarity,
overall the spectrum at this phase resembles that of normal SNIa.

\begin{figure*}
\psfig{figure=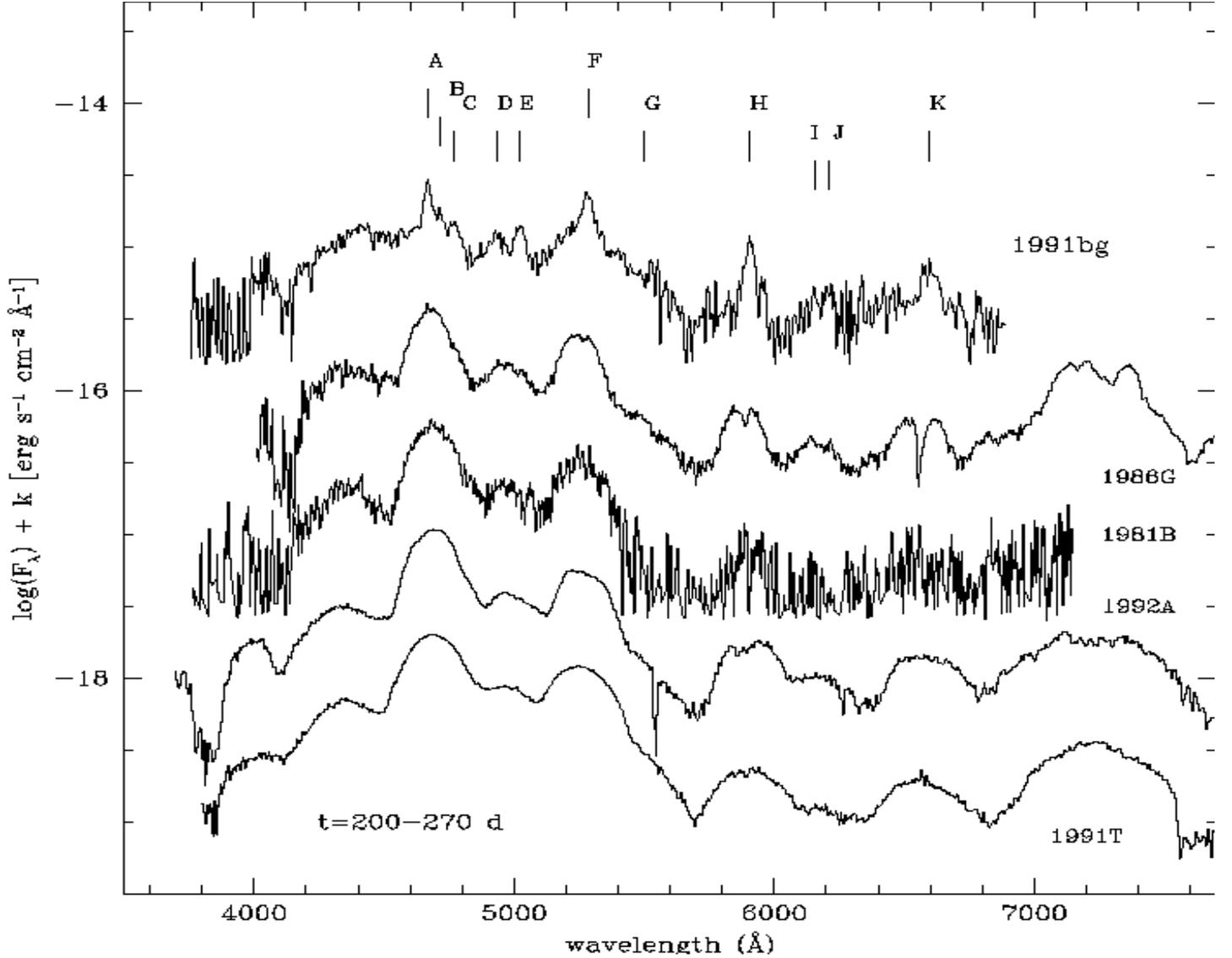,width=19cm,height=15cm}
\caption{Comparison of the last spectrum of SN 1991bg (203d past maximum)
with those of other SNIa at similar epochs: the normal SNe 1981B and
1992A, the peculiar SNe 1986G and 1991T.  The spectrum of SN~1986G has
been dereddened as in Fig.~8. The narrow emission lines
are marked according to Tab.~4.  The spectra have been displaced by
arbitrary units and reported to the rest frames of the parent
galaxies. The spectra of SNe 1981B and 1986G have the same sources as
in Fig.~8; the spectrum of SN~1992A has been obtained on Sept. 2, 1992
at the 3.6m telescope equipped with EFOSC; the same equipment was used
on Feb. 5, 1992 for SN~1991T.} \label{late}
\end{figure*}

\begin{table*}
\caption{Emissions in the nebular spectra of SN~1991bg} 
\label{narr}
\begin{tabular}{clccccc}
\hline
desig.&ident.&	& \multicolumn{4}{c}{rest frame position (\AA)}\\
\cline{4-7}
      & 	&epoch: & +88 d&+117 d&+143 d&+203 d\\
\hline
A	&[FeIII] 4658-4702	&&  	& 4665	&  4668	& 4670	\\
B	&[FeIII] 4733   	&&	&	&  4716 & 4714  \\
C	&[FeIII] 4755-4769-4778	&& 	& 	&  4765 & 4767  \\
D	&[FeIII] 4931		&&	& 4925	&  4928 & 4935  \\
E	&[FeIII] 5011		&&	& 5014	&  5027 & 5019  \\
F	&[FeII] 5261-5273, [FeIII] 5270	&&	& 5274	&  5280 & 5284  \\
G	&[FeII] 5527		&&  5542 & 5540  & 5536  & 5529  \\
H	&[CoIII] 5890-5908	&&  5906	& 5907	&  5908	& 5906  \\
I	&[CoIII] 6129		&&  6144:& 	&  6142	& 6157  \\
J	&[CoIII] 6197		&&  6204	& 6211	&  6212	& 6211  \\
K	&[CoIII] 6578, H$\alpha$ ?	&&  6584 & 6586 &  6592	& 6595  \\
\hline
\end{tabular}
\end{table*}

\subsection{The nebular phase} \label{nebu}

For the first time on day 34 we detect a relatively narrow emission
line redward of
the absorption at 5700 \AA, at about 5915 \AA\ (rest wavelength, 5897 \AA,
cfr. Fig.~\ref{sp_ev}). From this epoch on, and up to our last
observation, the relative intensity of this emission keeps increasing. 
The line profile, with a narrow core, shows very small or negligible
velocity evolution redwards. 
Because of the rest wavelength, the line was identified with the NaID
line \cite{fili,ruiz}, but in the light of the results of Mazzali et al. 
\shortcite{91bg:mod}
we propose an alternative identification with [CoIII] $\lambda
5890-5908$, although a contribution from NaID, especially at early
phases, cannot be ruled out.

The spectral evolution continues with the progressive decrease of the
continuum and the relative increase of the emission lines. The
absorptions fade, with the exception of the narrow feature at $\lambda
5700$.  With the third month the SN
enters the late decline phase of the light curve (cfr. Sect.~\ref{licu}).  
Corresponding with this transition 
the emission line of [CoIII] drifts by about 10 \AA\ to the
red, as measured by Leibundgut et al. \shortcite{leib}.
 
On day 117 the dominant feature is the [CaII] line at 7150-7400 \AA\
(cfr.  Fig.~\ref{earl}, bottom) which, as we mentioned, is much
stronger than in other SNIa at comparable epoch, while the blue part
of the spectrum shows the features typical of other SNIa, although
they are significantly narrower.  Also, the emissions at 4680 and 5290 \AA\
increase in relative strength as the SN ages.

The last available spectrum of SN~1991bg was 
obtained on July 3, 1992 (day 203). All lines present on days 117 and 143 
are visible and correspond to the typical emission lines in
the spectra of SNIa at this epoch (Fig.~\ref{late}).  In particular,
the bulk of the emission is between 3800 and 5700 \AA, with strong peaks
at about 4400, 4700, 5000 and 5300 \AA. All lines appears to be
significantly narrower than in other SNIa, confirming the indication 
that the expansion velocity in SN~1991bg was smaller than in typical SNe~Ia.

We note that our observations are in contradiction with those
of Ruiz--Lapuente et al. \shortcite{ruiz}.  While their May 24 1992,
spectrum resembles our spectrum of May 4,
their last spectrum (June 29) taken at the WHT appears remarkably
different from our spectra taken only 4 days later. In
particular, their spectrum does not show the strong emission features between
4500 and 5500 \AA, and is completely different from those of normal SNIa.

A late time spectrum of SN~1991bg, obtained at the same epoch and with
the same telescope, was shown by Gomez \& Lopez \shortcite{gome}. 
Although the authors do not state so,
this is probably a new reduction of the WHT observation reported by
Ruiz--Lapuente et al. \shortcite{ruiz}. Puzzling enough, the two
tracings do not show the same features. In particular, the narrow emissions 
seen in Ruiz--Lapuente et al. \shortcite{ruiz} near 5900 and 6570 \AA\ 
are absent in the Gomez \& Lopez \shortcite{gome} spectrum, which
does not reach wavelengths bluer than 4600 \AA.
Nevertheless also this spectrum is  definitely different from ours.

The most likely explanation for the disagreement between our spectrum
and those shown by Ruiz--Lapuente et al. \shortcite{ruiz} and Gomez \& Lopez
\shortcite{gome} is their inaccurate positioning of the slit on the target
and/or a non-alignment of the slit with the parallactic angle. 
In our case, exploiting the
multimode capability of EFOSC, the slit was accurately positioned
based on the previous imaging of the field in the B band.
An independent confirmation of our spectroscopic calibration is
obtained from the \bv\ color curve (Fig.~\ref{colo}) which indicates blue
colors for the SN at late epochs.

In Tab.~\ref{narr} we list the rest frame positions of the narrow
lines marked in the late time spectrum of Fig.~\ref{late}.  
Typical errors are $\pm1$ \AA\/ but for the features of
poorer signal--to--noise ratio for which the errors can be as large as
$\pm5$ \AA.
Line identifications are based on the NLTE synthetic nebular 
spectra presented by Mazzali et al. \shortcite{91bg:mod} 
which reflect the earlier results of 
Meyerott \shortcite{mey}. 
The identification of the $\lambda5906$ line with [CoIII] is supported
both by the line position and by the relative ratio of the flux with
the [FeIII] line at about 4700 \AA. In fact, according to Kuchner et
al. \shortcite{kuch} the time evolution of the Fe/Co flux ratio 
supports the idea that the Ni--Co--Fe chain powers SNIa. 
The value of this ratio in our nebular spectra are R$=2.5$, 3.0, 4.1
and 4.2 (with errors of the order of $20\%$) on day 88, 117, 143 and
203, respectively. These values are consistent with those of other
SNIa and with the Ni decay model for the origin of Co and Ni also in
this SN.

All the other emission features marked in Fig.~\ref{late} find
satisfactory identifications with forbidden lines of Fe--group
elements, with the exception of the feature labeled K, which, on our
spectra,  is
measured  at about $\lambda6590$\AA\ (rest wavelength)
with a FWHM$\sim2000$ \kms, similar to that of the [CoIII]
$\lambda5890-5908$ line.  This line is present on all spectra from day
79 onwards, and is also visible in the spectra by Leibundgut et
al. \shortcite{leib} starting at about the same epoch. The line was
present in the spectrum of Ruiz--Lapuente et al. \shortcite{ruiz} but
was measured at $\lambda6570$ and identified with
H$\alpha$. In their proposed scenario, 
the H$\alpha$ emission comes from
hydrogen  stripped from an
extended, hydrogen-rich companion star  as a consequence of the
explosion and  it remains at low velocity.
Given our measurement of the redshift of this feature of
1200 \kms with respect to the rest wavelength of H$\alpha$, we can
exclude the possibility that the line is due to H$\alpha$ emission
from hydrogen surrounding the exploding
star. However, we cannot rule out the possibility that the line arises
from hydrogen stripped from a companion which at the moment of the
explosion was on the far side of the SN with respect to the
observer. In this case, high velocity hydrogen blobs mixed with the
expanding material might emit redshifted lines.  However, since an
emission line is present at this wavelength in the nebular spectra of
all SNIa (cfr. Fig.~\ref{late}), and with a width comparable to that
of all other spectral features, if this line is attributed to hydrogen
in SN~1991bg a similar identification should be invoked for all other
SNIa together with the implication that this hydrogen is situated in
the expanding envelope.  We note that the possibility of 
interpreting this line as
H$\alpha$ was mentioned by Cristiani et al. \shortcite{cris} for
SN~1986G.  Another possible identification is with [CoIII]
$\lambda6578$, while lines of FeII and [FeII] also fall in this region
of the spectrum.  
Without modeling the identification with [CoIII] is not easy to
understand, since it appears to increase in strength relative to the
proposed [CoIII] 5890--5908 line. 
It should be further noted that this feature is
slightly redshifted in SN1991bg relative to the broader feature in
other supernovae, while there is no significant redward displacement
of other emission features at shorter wavelengths. Whether this points
to a different identification in this supernova from that in others
awaits clarification.

\section{Are there relatives of SN~1991\lowercase{bg} ?} \label{92k}

In recent years a number of SNe have been indicated as possible
relatives (ranging from ``twins'' to ``close cousins'') of SN~1991bg.

We already mentioned that SN~1986G was found to share some of the
properties of SN~1991bg (cfr Sect.
\ref{pe} \& \ref{spev}), in particular the
peculiar broad absorption between 4200 and 4500 \AA.  Also, the early
light curve decline was fast but the duration and the
slopes of the various portions of the light curve were different
\cite{leib} and so was the \bv\ color evolution. 
The absolute magnitude of SN~1986G may have been similar to that of a 
normal SNIa, but because of the conflicting evidence on the reddening
for SN~1986G (cfr. Sect.~5 of Filippenko et al.  1992) no conclusive 
statement can be made. The late time spectra of SN~1986G showed
relatively strong [CaII] lines, similar to SN~1991bg (cfr.
Fig.~\ref{earl}, bottom panel), and also the lines were somewhat
narrower than in normal SNIa (cfr. Tab.~\ref{explate}) but still
broader than in SN~1991bg. It appears, therefore, that
while the chemical composition and the physical conditions of the
outer layers were rather similar in the two SNe, both showing red
continua with TiII lines, at least the kinematics of the
inner layers were different.

Filippenko et al. \shortcite{fili} suggested that SN~1971I is another
possible relative of SN~1991bg.  Inspection of the light curve
shows that this SN, though it faded relatively fast (12.6 \m100 in B),
maintained a close resemblance to typical SNIa until day 300.
Moreover, the 4200--4500 \AA\ absorption band is absent in the
spectra taken at maximum \cite{bcr2,kiku} and the
SiII line indicate a normal expansion velocity (about 11600 \kms\ a
couple of days before maximum).

\begin{figure}
\psfig{figure=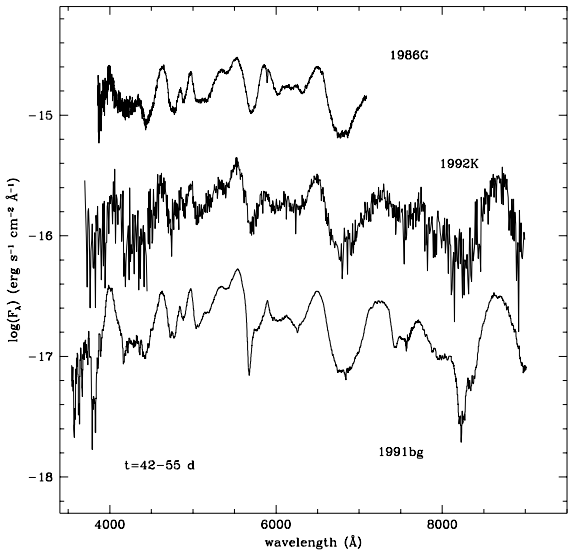,width=8.5cm}
\caption{Comparison of the spectrum of SN 1992K taken on April 7 1992
at the 2.2m telescope (with EFOSC2 and grisms n.~1, 5 and 6) and
corresponding to an epoch t=+42d (1991bg templates), with those of SN
1991bg on day +46 and SN 1986G on day +55 (Cristiani et al. 1992). The
spectrum of SN~1986G has been dereddened as in Fig.~8 and 9.
The ordinates are relative to the spectrum of SN~1992K, while the
other two have been shifted by arbitrary units for an easier
comparison. In abscissa are the rest frame wavelengths.}
\label{92k_fig} 
\end{figure}

In a recent paper by Hamuy et al. \shortcite{hamu} it has been
proposed that SN 1992K in ESO 269-G57 is a twin of SN~1991bg. 
This SN has the broad absorption between 4200 and 4500 \AA, a somewhat slow
expansion velocity, an intrinsic red continuum and a low luminosity.  In
Fig.~\ref{92k_fig} we show the comparison of a spectrum of SN 1992K
taken at La Silla on April 7, 1992, with those of SNe 1991bg and 1986G
at a similar age.  The three spectra are rather similar, both in the
blue, with the 4200--4500 \AA\ trough, and in the near IR, with the
presence of strong Ca IR triplet and [CaII] $\lambda7291, 7324$.
In SN~1992K, however, the features at 5700 and 5900 \AA\
are broader and similar to SN~1986G.

Although SN~1992K was not observed at maximum, the comparison of the light
curves with those of five different  templates of SNIa has shown that
SN~1992K was indeed very similar to SN~1991bg \cite{hamu}. In particular,
between day 70 and 150, SN~1992K had a decline rate, $\gamma_V=2.75$
\m100, which is identical to that of SN~1991bg (cfr. Tab.~\ref{data}).
Based on this similarity it was possible to estimate that the maximum of
SN~1992K occurred probably on Feb. 25, 1992, at an absolute magnitude
\Mb$=-16.84$, slightly brighter (0.4 mag) than 1991bg but still
fainter than normal SNIa (we note that also for SN~1992K there is  
no evidence of significant reddening).



Another SN that has been proposed for inclusion in the SN~1991bg family 
is SN~1991F in NGC~3458 \cite{gome}. This SN was observed
only spectroscopically at epochs estimated between 3 and 5 months past
maximum.  Similar to SN~1991bg at the same epoch, the spectrum of
SN~1991F is dominated by emission lines which are narrower than in
normal SNIa, indicating a lower expansion velocity; and it shows
strong CaII lines. The presence, in the
spectrum taken 80-90 days past maximum, of two very narrow features in
absorption at $\lambda5700$ and in emission at $\lambda5900$, with profiles 
very similar to those of the corresponding features in SN~1991bg, is 
of interest. However, the position of the emission
in SN~1991bg (Tab.~\ref{narr}) is about 20-30 \AA\ redder than in
SN~1991F (Tab.~2 of Gomez \& Lopez 1995) at the same
epochs, and, unlike SN~1991bg, the $\lambda5700$ absorption in 
SN~1991F weakens
with age.  Moreover, the [FeII]+[FeIII] emission at about 5200 \AA\ is
still weak at t$=+144$, while in SN~1991bg at the same epoch it is 
prominent.
Rough estimates based on spectrophotometry \cite{gome} seem to indicate 
that the late-time luminosity and the decline rate of SN~1991F are comparable 
to those of SN~1991bg.

In summary, it appears that there are at least three SNIa (namely SNe
1986G, 1991F and 1992K) which share some of the peculiar
characteristics of SN~1991bg, in particular the low luminosity, fast
fading rate, red colour at maximum, low expansion velocities, evidence
of TiII lines in the photospheric spectrum, strong CaII and narrow
emission lines at late epochs.  However, none of these SNe displayed 
all these features simultaneously and with such strength as SN~1991bg.

The relative frequency of objects such as 1991bg is difficult to estimate.
Because SN searches are magnitude limited, the
discovery probability of intrinsically faint SN~Ia is  lower than
that of normal Ia. Therefore, the fraction of faint SN~Ia so far discovered
(about 5\%) must be considered as a lower limit of the true value. On
the other hand, the relative rate of faint SN~Ia can be estimated
based on the few observed events by using the control time method if
the details of individual SN searches were known. At
present this kind of data is available only for the Asiago+Crimea SN
search \cite{cap}. From this database we estimate that faint
(1991bg--like) SN~Ia are
at most 40\% of all SN~Ia, hence the real number should be between 5
and 40\%.

\section{Discussion} \label{disc}

According to the commonly accepted classification criteria, SN~1991bg was
certainly of type Ia. The early spectrum shows most of the typical SN~Ia 
features due to intermediate mass elements, in particular the strong 
SiII $\lambda6355$ absorption (Fig.~\ref{earl}), although with noticeable 
differences in the relative line intensities.  This implies
that the physical conditions 
in the near-photospheric layers around maximum light and the outer layers of 
the exploding star were not dramatically different from the usual ones.

Contrary to previous claims, we have shown in Sect.~\ref{nebu} that
also in the nebular spectra of SN~1991bg, in particular in the latest
observation at t$=203$d (Fig.~\ref{late}), the relative intensities of
the emissions match those of normal SNIa.  Since at late epochs the
line forming region is deep inside the ejecta,
this indicates that the composition as well as the physical conditions
of the interior (temperature, density, ionization, etc.) are similar too.
However, the lines are significantly narrower, hence the
expansion velocity of the ejecta must be smaller.  
The luminosity at maximum (\Mb$=-16.54$) is about 2.5 magnitudes
fainter than normal SNIa (cfr. Sect.~\ref{abma}) while 
the bolometric luminosity is about 0.5 dex fainter than SN~1992A.


It is believed that the light curve of SNe is powered by the
radioactive decay chain $^{56}$Ni--$^{56}$Co--$^{56}$Fe and that, to a first
approximation, the peak (bolometric) luminosity is equal to the
instantaneous emission of the radioactive decay
\cite{arne}. As a consequence there is a direct relation between the mass of
synthesized $^{56}$Ni and the luminosity of the SN: since most SNIa
show only a small scatter in absolute magnitude, this implies that
most SNIa produce similar amounts of radioactive Ni.  At present,
explosion models suggest that normal SNIa produce 0.6M$_\odot$ of Ni
(e.g. Nomoto et al. \shortcite{nomo84}.  

Therefore if we accept that a normal SNIa produces approximately 0.6~M$_\odot$
of Ni and, from the discussion above, that SN 1991bg had a bolometric
luminosity 0.5 dex fainter than SN 1992A, we conclude
that SN 1991bg produced $<0.2$ M$_\odot$ of Ni (in Sect. 3.4 we noted
that SN~1992A might be fainter than normal SNIa). This mass could
conceivably be lower still depending on the nature of the explosion
and of course on whether SN 1992A produced a smaller mass of Ni. For
example H\"{o}flich et al. \shortcite{hofl+khok} have demonstrated
with a sample of different explosion models that the factor involved
in relating bolometric luminosity at maximum to the $\gamma$-ray
luminosity and hence to mass of Ni could vary by as much as
approximately 50 percent. A slightly larger difference in bolometric
luminosity between the above two SNe is apparent at 100 days. However
an added complication in making a comparative quantitative estimate of
Ni mass is the as yet unknown possible difference in $\gamma$-ray
opacity of the two envelopes also at these later phases.

The mass of $^{56}$Ni has been used has an input parameter also in the
computation of the late--time synthetic spectra by Mazzali et al.  
\shortcite{91bg:mod}.
Consistently low values of M$(^{56}Ni)$ are also suggested in
order to reproduce the observed fluxes and line widths in that work. 

The rapid luminosity decline of SN~1991bg at early epochs is unique,
and is indicative of a small ejecta mass  \cite{fili,leib}.
If the envelope is small, the diffusion time is short
and the trapping of the $\gamma$-rays from the radioactive decay of
$^{56}Ni$ is less efficient. Therefore, the envelope cools more rapidly,
the photosphere recedes faster in mass coordinates and the luminosity
decline is faster.


With the noticeable exception of few peculiar SNe (e.g. SN~1988Z,
Turatto et al. \shortcite{tur:88z}), the late-time light curve of SNe
is powered by the radioactive decay 
of $^{56}$Co into $^{56}$Fe, with an $e$-folding time of 111 days. The
decline rate actually observed depends on the fraction of energy
deposited. In the massive SNII the envelope is for at least two years 
optically thick to the hard radiation originating in the decay and the
luminosity decline rate, $\sim 1.0$ \m100, closely matches the
radioactive input. Instead, in SNIa the envelope becomes progressively
more transparent to the $\gamma$-rays and the observed luminosity
decline is steeper than the radioactive energy release.  In normal
SNIa a decline rate of 1.5 \m100 is observed \cite{tur:90}.  The very
fast decline rate observed in SN~1991bg (2.5 \m100) is probably due to
the small envelope mass.

Eventually, the envelope becomes transparent to $\gamma$-rays and only
the kinetic energy of the positron (about 5\% of the total decay energy) 
is deposited. At this point, the light curve is expected to
approach the $^{56}$Co decline rate.  Indeed, a significant
flattening of the light curve is indicated by the last observations of
SN~1991bg. The observations available allow only a setting of an upper
limit to the late decline rate ($\le 1.2$ \m100). This is consistent
with the $^{56}$Co input.  

Few other SNIa  have been observed at phases later than 500 days.  One of
these was SN~1937C. This SN, although showing a flattening in the
photographic light curve around day 300, had a steeper
decline rate (1.31 \m100) until day 600 \cite{scha}. The case of SN~1992A 
is different. For this SN recent HST observations indicate a
flattening of the light curve after day 600 to a rate slower than the
$^{56}$Co decay rate, suggesting additional energy
sources such as $^{44}$Ti decay, ionization freeze--out or light echoes. 
In the case of SN 1991T observed at late phases, a light echo began to 
influence the light curve and caused it to decline considerably slower 
than would be expected from $^{56}$Co decay \cite{schm}. This possibility 
for SN 1991bg cannot be excluded by current observations. 

The main spectral peculiarities of SN 1991bg in the photospheric epoch
are the low photospheric expansion velocity, the red colour and the
presence of the absorption band between 4200 and 4500 \AA.  It is
shown by Mazzali et al. \shortcite{91bg:mod} that these features can
be satisfactorily reproduced if the SN is underluminous and if the
abundance of the Fe--group elements above 3500 \kms is low compared to
the standard W7 model.  An overabundance of Ti with respect to other
intermediate mass elements is not required in order to explain the
$\lambda$4200--4500 feature, which is the result of the low ionization
due to the low luminosity.

The peculiarities of the late time spectra of SN~1991bg with respect
to other SNIa are highlighted in Fig.~\ref{late}. The SNe are plotted
from top to bottom in sequence of increasing line widths.  It is
interesting to note that the objects define also a sequence of
absolute magnitudes. In Table~\ref{explate} we list the absolute
magnitudes at maximum, the value of $\Delta m_{15}$(B), the
photospheric velocity at maximum as derived from the SiII line (which
may be an overestimate of the actual value, cfr. Patat et al 1995),
and the FWHM of the [CoIII] lines in the late time spectra between day
200 and 270. The [CoIII] line has been measured because it may be less
affected by blending than other lines, e.g. [FeII] and [FeIII], but
the same trend holds also for other emission lines in the nebular
spectrum. 

A relation between the peak luminosity and the rate of decline, with
faster fading SNIa being fainter, was first suggested by Pskovskii
\shortcite{psko1} and  has recently been confirmed on the basis of 
accurate CCD photometry \cite{phil,hamuh0}.  The trend is well
illustrated by the SNe in Table~\ref{explate}, which are a subsample
of those of Phillips \shortcite{phil}.  From 
Table~\ref{explate} we also note that the correlation extends also to the
photospheric expansion velocities at maximum and to the expansion
velocities of the innermost layers as deduced from the line widths
of the Fe-peak elements.
The very small velocity in SN~1991bg suggests that in
this SN complete nuclear burning was confined to the innermost
regions. In other words,  observations indicate that SNIa with more slowly
expanding ejecta, in particular SN~1991bg, have a more rapid
photometric evolution and reach a fainter absolute magnitude than
fast expanding objects.

Arnett \shortcite{arn82} has shown how the $^{56}$Ni mass is directly
proportional to the square of the velocity scale of expansion. Both
the low photosphere and nebular velocities therefore point directly
to a low mass of Ni produced in SN 1991bg. Nevertheless the
quantification of this finding is beset by difficulties in defining
observationally which velocity should be used. At early phases, because
of the great strength of the SiII line, a photospheric velocity is not
well defined, while at the late phases the velocities from the widths
of the emission lines refer to material concentrated in the innermost
parts of the exploding star. 

\begin{table}
\caption{Expansion velocities vs. absolute magnitude
for SNIa}
\label{explate}
\begin{tabular}{llcccc}
\hline
SN  & M$_V^{max}$&$\Delta m_{15}(B)$&  v(SiII)$^{\bf a}$  &
FWHM([CoIII])$^{\bf b}$\\
    &		&    & [\kms] & [\kms][day] \\
\hline
1991bg & -17.28 & 1.95 & 9800 & 2340~[203]\\  
1986G  & -18.22$^{\bf c}$ & 1.73 &	10400 &7120~[256] \\ 
1981B  & -18.50 & 1.10 & 11900 & 8140:~[270]\\ 
1992A  & -18.00 & 1.33 & 11700 & 10170~[227] \\ 
1991T  & -19.51$^{\bf d}$ & 0.94 &	9800$^{\bf e}$ & 11190~[262] \\ 
\hline
\end{tabular}
(a) photospheric velocity at B maximum as measured from the SiII
$\lambda6355$ absorption; (b) Full Width Half Maximum of [CoIII] $\lambda
5890-5908$ emission at epochs between 200 and 270 days
(indicated in brackets); (c) E(B-V)$=0.6$ (Phillips 1993); (d)
E(B-V)$=0.13$ (Phillips et al.  1992); (e) the expansion velocity of
SiII in SN~1991T was a poor indicator of the photospheric velocity.
Velocities measured from other lines were significantly larger than 
in normal SNIa (Phillips et al. 1992).
\end{table}

A number of possible explosion scenarios have been proposed to explain
the peculiar characteristics of SN~1991bg, and in general the
faint SNIa.

Hoflich et al. \shortcite{hofl} used Delayed Detonation
models to explain SNIa light curves. They argued that, depending on
the density at which the transition from deflagration to detonation
occurs, different $^{56}$Ni masses and, therefore, SNIa of different
luminosity can be produced by the same explosion mechanism in a
Chandrasekhar mass C-O WD. In particular, their PDD5 model, which
produces 0.12 M$_\odot$ of $^{56}$Ni, gives the best fit to the SN 1991bg
observations. In this model the $^{56}$Ni is strongly concentrated in
the center, as indicated also by our observations, and a major fraction of
the mass is in intermediate mass elements, with only about 0.05 M$_\odot$ of
unburned C/O left after the explosion.
However, this model requires a large reddening, E(B-V)=0.30, in order
to reconcile the predicted luminosity with the observed value, while
observations indicate a much smaller value, E(B-V)$=0.05$ (cfr.
Sect. \ref{abma}). Also, the decline rates of the PDD5 model are
similar to those of their standard Ia model (N32), while we found SN
1991bg to be faster (cfr. Sect. \ref{licu}).


An alternative scenario for faint SNIa was proposed by Livne
\shortcite{livn} and Woosley \& Weaver \shortcite{woos}. In their models
a C--O white dwarf with mass below the Chandrasekhar limit
accretes He at a rate of the order of a few $10^{-8}$ M$_\odot$
yr$^{-1}$ until He ignites at the bottom of the layer. 
The in-going shock wave 
induces a detonation of the C-O which disrupts the star. According to Woosley
\& Weaver \shortcite{woos} less massive white dwarfs produce smaller
$^{56}$Ni masses and less energy, in qualitative agreement with our
findings.  The lower mass WD results in a larger fraction of
intermediate mass elements and an overproduction of other isotopes
(including Ti$^{44}$) in the He detonation layer, while a
large fraction of He is left unburned.   We already noted, however,
that an overproduction of $^{44}$Ti is not required by the spectral
synthesis models. 
The visibility of He in the ejecta is not expected because of the
observed low temperatures. 
Moreover, the models of Mazzali et al. \shortcite{91bg:mod} seem to
exclude the short rise times to maximum light (of the order of 13d)
which are obtained by Woosley \& Weaver \shortcite{woos}.  These
sub--Chandrasekhar models produce in the interior a nearly constant
velocity as a function of mass and, therefore, they do not the
reproduce the low velocities observed in SN~1991bg at late stages.

In another explosion mechanism, proposed by Nomoto et al. \shortcite{nomo}, 
a low mass of $^{56}$Ni is produced by the collapse of a
O--Ne--Mg WD formed after the merging of a double C--O WD system.  In
order to reproduce the observed luminosity the collapsing core has to
be embedded in a C--O envelope of 0.6 M$_\odot$ and the shock wave
propagating through it produces 0.15 M$_\odot$ of $^{56}$Ni, 
compatible with the estimates found above.  The
resulting light curve reproduces satisfactorily the observed one in
the early 100 days but Si seems to be located at velocities too high
to reproduce the observed lines.

Unfortunately the observations alone do not suggest unambiguously a
preferred model for the progenitor. All suffer from the various
difficulties discussed above. Mazzali et al. \shortcite{91bg:mod} show
through modelling that even in the most promising models one has to
resort to ad hoc adjustments of abundances and their stratification to
approximate the observed spectra. Thus SN 1991bg has provided an
additional challenge to our understanding the Type Ia supernova
phenomena. 

\section{Conclusions}

The new observations presented in this paper confirm the
peculiar characteristics of SN~1991bg and give new insights into its
late time behaviour.

In particular, SN~1991bg was fainter than normal SNIa by about 2.5 magnitudes 
in the B band, but by less than 2 mags in the V band, thus giving a
particularly red colour at maximum, (B-V)$_{max}=0.74$.  However,
because of a different colour evolution, the SN turned to the colour curve
of normal SNIa already 40--50 days later.  The rates of decline in all
optical bands (BVRI) are at all epochs the most rapid ever observed in
SNIa (Tab.~\ref{data}).  The constructed {\em uvoir} bolometric light curve 
is consequently steeper than in the normal SN~Ia SN~1992A.  
A very deep V band observation obtained 530 days past maximum showed
that before this epoch the fading rate had decreased to about 1.2
\m100, and we stress that, on the basis of the available data,
energy sources other than Co decay are not required.

The low luminosity at maximum is an indication that the SN produced a
small mass of Ni ($<0.2$ M$_\odot$), whereas the fast photometric
evolution was related to a small mass of the envelope.  The small
explosion energy also causes the photosphere at maximum light to be
cooler than in typical SNIa, as is evident when comparing the early
time spectra.  This also explains the broad absorption feature between
4200-4500 \AA\ identified with TiII, which has been attributed to an
ionization effect by Mazzali et al. \shortcite{91bg:mod}.

A peculiarity of the spectrum of SN~1991bg at epochs subsequent to maximum 
is the presence of the two narrow features at about 5700\AA\ (in absorption)
and 5900\AA\ (in emission). The absorption lacks at the moment a
convincing explanation mainly because there is no significant
evolution redwards, 
while the emission, which appeared already on day 34, is plausibly an
early emergence of [CoIII] in accordance with the fast evolution to the
nebular stage shown by SN~1991bg.  It has been noted that the
spectrum of SN~1991bg evolved to the nebular stage earlier than usual.


Contrary to previous claims, the overall appearance of the spectrum
maintained a general resemblance to those of other SNIa at least until
200 days after maximum light (Fig.~\ref{late}), indicating that the
ionization conditions of the Fe core were similar to those of other
SNIa at this phase. In particular, there has been no sudden change in
the nature of the spectrum during the interval day 143 to day 203.  
Nevertheless, the emission lines were
exceptionally narrow, and allowed the identification of the main
emissions with lines of [FeII], [FeIII] and [CoIII].  The presence of
hydrogen at late times seems unlikely.

The small photospheric expansion velocity and envelope mass, 
combined with the the small expansion velocity and mass of the
Fe--group core, as seen from the late-time spectra and luminosity, point to
a kinetic energy smaller than in typical SNIa.  We pointed out that
the correlation between the luminosity at maximum and the early rate
of decline can be extended also to the expansion velocities of the photosphere
at maximum and to the innermost layers emitting the nebular spectrum
(Tab. \ref{explate}). In other words, the luminosity at maximum
correlates to the kinetic energy of the SNIa.

One important question is whether the lower energy SN~1991bg can be
considered an extreme case of a continuous distribution of SNIa or
whether it is a representative of a separate subclass of faint
SNe~Ia. We reviewed the literature and found that while a handful of
other SNIa shared some of the observed characteristics of SN~1991bg,
none was quite so extreme. This may support the concept of a continuum
transition from faint SNIa (SN~1991bg--like) to {\em normal}
ones. Obviously, a continuum of SN~Ia properties might jeopardize
their use as standard candles, 
at least until one understands better the inter-relation between light
curves and spectral characteristics.

\bigskip

\noindent
{\bf ACKNOWLEDGMENTS} We thank Roberto Rampazzo and Caterina Zanin
for obtaining some of the observations reported in this paper.

This work has been conducted as part of the ESO Key Programme on
Supernovae


\noindent


\begin{thebibliography}{cosaservira}     

\bibitem[\protect\citename{Arnett }1982]{arn82}
	Arnett, W.D., 1982, in ``Supernovae: A Survey of Current 
	Research'', D.~Reidel: Dordrecht, Holland, p.~221 
\bibitem[\protect\citename{Arnett et al. }1985]{arne}
        Arnett,W.D., Branch,D., Wheeler,J.C., 1985, Nature 314 337
\bibitem[\protect\citename{Barbon et al. }1973]{bcr}
        Barbon,R., Ciatti,F., Rosino,L., 1973, A\&A 25, 241
\bibitem[\protect\citename{Barbon et al. }1973]{bcr2}
        Barbon,R., Ciatti,F., Rosino,L., 1973, Mem.S.A.It. 44, 65
\bibitem[\protect\citename{Barbon et al. }1990]{barb}
        Barbon,R., Benetti,S., Cappellaro,E., Rosino,L., Turatto,M., 
	1990, A\&A 237, 79
\bibitem[\protect\citename{Benetti et al. }1991]{ben91}
        Benetti,S., Cappellaro,E., Turatto,M., 1991, IAUC 5405
\bibitem[\protect\citename{Branch }1984]{bran}
	Branch,D., 1984, in Evans D.S., ed., XI Texas Symposium on
	Relativistic Astrophysics, N.Y. Academy of Science,  422, 186
\bibitem[\protect\citename{Branch }1992]{bran_h0}
	Branch,D., 1992, ApJ 392, 35 
\bibitem[\protect\citename{Burstein \& Heiles }1984]{burs}
        Burstein,D., Heiles,C., 1984, ApJS 54, 33
\bibitem[\protect\citename{Cappellaro et al. }1993]{cap}
        Cappellaro, E., Turatto, M., Benetti, S., Tsvetkov, D.Yu., 
	Bartunov, O.S., Makarova, I.N., 1993, A\&A 273, 383 
\bibitem[\protect\citename{Cappellaro et al. }1995]{ulda}
        Cappellaro,E., Turatto,M., Fernley,J., 1995, Supernovae, IUE -- ULDA 
	Access 	Guide N.6, ESA SP-1189, Noordwijk 
\bibitem[\protect\citename{Ciardullo et al. }1993]{ciar}
        Ciardullo,R., Jacoby,G.H., Tonry,J.L., 1993, ApJ 419, 479
\bibitem[\protect\citename{Cristiani et al. }1992]{cris}
        Cristiani,S., Cappellaro E., Turatto M., Bergeron,J., Bues,I.,
	Buson,L.M., Danziger,I.J., Di Serego Alighieri,S., Duerbeck,H.W.,
	Heydari-Malayeri,M., Schmutz,W., Schulte-Ladbeck,R.E.,  
	1992. A\&A 259, 63
\bibitem[\protect\citename{Filippenko et al. }1992]{fili}
	Filippenko,A.V., Richmond,M.W., Branch,D. Gaskel,C.M.,
	Herbst,W., Ford,C.H., Treffers,R.R., Matheson,T., Ho,L.C., Dey,A.,
	Sargent,W.L.W., Small,T.A., Bruegel,W.J.M.,  1992, AJ 104, 1543.
\bibitem[\protect\citename{Frogel et al. }1987]{frog}
	Frogel,J.A., Gregory,B., Kawara,K., Phillips,M.M., Laney,D., 1987,
	ApJ 315, L129
\bibitem[\protect\citename{Gomez \& Lopez }1995]{gome}
	Gomez,G., Lopez,R., 1995, AJ 109, 737
\bibitem[\protect\citename{Hamuy et al. }1994]{hamu}
	Hamuy,M., Phillips,M.M., Maza,J. Suntzeff,N.B., 
	Della Valle,M., Danziger,I.J., Antezana,R., Wischnjwesky,M.,
	Aviles,R., Schommer,R.A., Kim,Y.C., Wells,L.A., Ruiz,M.T.,
	Prosser,C.F., Krzeminski,W., Baylin,C.D., Hartigan,P., Hughes,J., 
	1994, AJ 108, 2226
\bibitem[\protect\citename{Hamuy et al. }1995]{hamuh0}
	Hamuy,M., Phillips,M.M., Maza,J., Suntzeff,N.B., Schommer,R.A.,
	Aviles,R., 1995, AJ 109, 1
\bibitem[\protect\citename{Hoflich et al. }1995]{hofl}
        Hoflich,P., Khokhlov,A.M., Wheeler,J.C., 1995, ApJ 444, 831
\bibitem[\protect\citename{Hoflich \& Kochlov}1995]{hofl+khok}
        Hoflich,P., Khokhlov,A.M., 1996, ApJ 457, 500
\bibitem[\protect\citename{Kikuchi }1971]{kiku}
        Kikuchi,S., 1971, PASJ 23, 593
\bibitem[\protect\citename{Kosai et al. }1991]{kosa}
	Kosai,H, Kushida,R., Kushida,H., Kato,T., 1991, IAUC 5400
\bibitem[\protect\citename{Kuchner et al. }1994]{kuch}
        Kuchner,M.J., Kirshner,R.P., Pinto,P.A., Leibundgut,B., 1994, ApJ
	426, L89 
\bibitem[\protect\citename{Landolt }1992]{land}
	Landolt A.U., 1992, AJ 104, 340
\bibitem[\protect\citename{Leibundgut et al. }1993]{leib}
 	Leibundgut,B., Kirshner,R., Phillips,M.M., et al., 1993, AJ 105, 301
\bibitem[\protect\citename{Livne }1990]{livn}
        Livne,E., 1990, ApJ 354, L53
\bibitem[\protect\citename{Maza et al. }1994]{maza}
        Maza,J., Hamuy,M., Phillips,M.M., Suntzeff,N., Aviles,R.,
	1994, ApJ 424, L107
\bibitem[\protect\citename{Mazzali et al. }1995]{mazz95}
	Mazzali P.A., Danziger,I.J., Turatto,M., 1995, A\&A 297, 509
\bibitem[\protect\citename{Mazzali et al. }1996]{91bg:mod}
        Mazzali P.A., Chugai, N., Turatto, M., Lucy, L., Danziger,
	I.J., Cappellaro, E., Della Valle, M., Benetti, S.., 1996,
	MNRAS submitted
\bibitem[\protect\citename{Meyerott }1980]{mey}
        Meyerott, R.E., 1980, ApJ 239, 257         
\bibitem[\protect\citename{Nomoto et al. }1984]{nomo84}
	Nomoto,K., Thielemann, F.K., Yokoi, K., 1984, ApJ 286, 644
\bibitem[\protect\citename{Nomoto et al. }1995]{nomo}
	Nomoto,K., Yamaoka,H., Shigeyama,T., Iwamoto,K.,
	1995, in McCray,R. \& Wang,Z., ed., 
	Supernova and Supernova
	Remnants, Cambridge University press, Cambridge, in press
\bibitem[\protect\citename{Patat et al. }1996]{pata}
        Patat,F., Benetti,S., Cappellaro,E., Danziger,I.J., Della Valle,M.,
	Mazzali,P., Turatto,M., 1996, MNRAS 278, 111
\bibitem[\protect\citename{Porter et al. }1992]{port}
        Porter,A.C., Dickinson,M., Stanford,S.A., Lada,E.A., 
	Fuller,G.A., 1992, BAAS 24, 1244
\bibitem[\protect\citename{Phillips et al. }1992]{phil_91t}
	Phillips,M.M., Well,L.A., Suntzeff,N.B., Hamuy,M., 
	Leibundgut,B., Kirshner,R.P., Foltz,C.B., 1992, AJ 103, 1632
\bibitem[\protect\citename{Phillips }1993]{phil}
        Phillips,M.M., 1993, ApJ 413, L105
\bibitem[\protect\citename{Pskovskii }1967]{psko1}
	Pskovskii,Y.P., 1967, SA 11, 63
\bibitem[\protect\citename{Pskovskii }1971]{psko2}
	Pskovskii,Y.P., 1971, SA 14, 798
\bibitem[\protect\citename{Ruiz-Lapuente et al. }1992]{ruiz_91t}
	Ruiz-Lapuente,P., Cappellaro,E., Turatto,M., Gouiffes,C.,
	Danziger,I.J., Della Valle,M., Lucy,L.B., 1992, ApJ 387, L33
\bibitem[\protect\citename{Ruiz-Lapuente et al. }1993]{ruiz}
	Ruiz-Lapuente,P., Jeffery,D.J., Challis,P.M., Filippenko,A.V.,
	Kirshner,R.P., Ho,L.C., Schmidt,B.P., Sanchez,F., Canal,R.,
	1993, Nature 365, 728
\bibitem[\protect\citename{Saha et al. }1995]{saha}
        Saha,A., Sandage,A., Labhardt,L., Schwengeler,H.,
        Tammann,G.A., Panagia,N., Macchetto,F.D., 1995, Ap. J. 438, 8
\bibitem[\protect\citename{Sandage \& Tammann }1993]{sata}
	Sandage,A., Tammann,G.A., 1993, ApJ 415, 1
\bibitem[\protect\citename{Schaeffer }1994]{scha}
	Schaeffer,B.E., 1994, ApJ 426, 493
\bibitem[\protect\citename{Schmidt et al. }1995]{schm}
	Schmidt,B.P., Kirshner,R.P., Leibundgut,B., 
	Wells,L.A., Porter,A.C., Ruiz--Lapuente,P., 
	Challis,P., Filippenko,A.V., 1995, ApJ 434, L19
\bibitem[\protect\citename{Suntzeff }1995]{sunt}
        Suntzeff,N.B., 1995, in McCray,R. \& Wang,Z., ed., 
	Supernova and Supernova
	Remnants, Cambridge University press, Cambridge, in press
\bibitem[\protect\citename{Tully }1988]{tully}
	Tully, R.B., 1988,  Nearby Galaxy Catalog, Cambridge University
	Press, Cambridge
\bibitem[\protect\citename{Turatto et al. }1990]{tur:90}
	Turatto M., Cappellaro E., Barbon R., Della Valle M., 
	Ortolani S., Rosino L., 1990, AJ, 100, 771
\bibitem[\protect\citename{Turatto et al. }1993a]{tur:88a}
	Turatto M., Cappellaro E., Benetti,S., Danziger I.J., 1993, 
	MNRAS, 265, 471
\bibitem[\protect\citename{Turatto et al. }1993b]{tur:88z}
	Turatto M., Cappellaro E.,  Danziger I.J., Benetti,S.,
	Gouiffes, C., Della Valle,M., 1993, 
	MNRAS, 262, 128
\bibitem[\protect\citename{Turatto et al. }1994]{tur:e}
        Turatto M., Cappellaro E., Benetti,S., 1994, AJ 108, 202
\bibitem[\protect\citename{Vaughan et al. }1995]{vaug}
        Vaughan,T.E., Branch,D., Miller,L., Perlmutter,S., 1995, ApJ
	439, 558
\bibitem[\protect\citename{Weaver et al. }1980]{weav}
	Weaver,T.A., Axelrod,T.S., Woosley,S.E., in Wheeler,J.C., ed.,
        Proc. Workshop on Type I Supernovae, University of Texas,
	Austin, p.113
\bibitem[\protect\citename{Wheeler \& Benetti }1995]{whee}
	Wheeler,J.C., Benetti,S., 1995,  Astrophysical 
	Quantities, IV$^{th}$ edition, ed. Arthur N. Cox, in press
\bibitem[\protect\citename{Woosley \& Weaver}1994]{woos}
        Woosley,S.E., Weaver,T.A., 1994, ApJ 423, 371

\end{thebibliography}
\end{document}